\RequirePackage{lineno}
\documentclass[twocolumn,aps,prc,showpacs,superscriptaddress,floatfix]{revtex4-1}
\usepackage{rotating}
\usepackage{epsfig}
\usepackage{hyperref}
\usepackage{lineno}
\usepackage{url}

\usepackage{amsmath}
\usepackage{graphicx}
\usepackage[caption=false]{subfig}

\DeclareGraphicsExtensions{ .pdf, .png}

\makeatletter
\renewcommand{\p@subsection}{}
\renewcommand{\p@subsubsection}{}
\makeatother

\begin{document}
\title{Responses of the chiral-magnetic-effect-sensitive sine observable to resonance backgrounds in heavy-ion collisions}

\author{Yicheng Feng}
\email{feng216@purdue.edu}
\address{Department of Physics and Astronomy, Purdue University, West Lafayette, IN 47907, USA}

\author{Jie Zhao}
\email{zhao656@purdue.edu}
\address{Department of Physics and Astronomy, Purdue University, West Lafayette, IN 47907, USA}

\author{Fuqiang Wang}
\email{fqwang@purdue.edu}
\address{Department of Physics and Astronomy, Purdue University, West Lafayette, IN 47907, USA}
\address{School of Science, Huzhou University, Huzhou, Zhejiang 313000, China}

\date{\today} 


\begin{abstract}

A new sine observable, $R_{\Psi_2}(\Delta S)$, has been proposed to measure the chiral magnetic effect (CME) in heavy-ion collisions; 
$\Delta S = \left \langle \sin \varphi_+ \right \rangle - \left \langle \sin \varphi_- \right \rangle$, where $\varphi_\pm$ are azimuthal angles of positively and negatively charged particles relative to the reaction plane and averages are event-wise, and $R_{\Psi_2}(\Delta S)$ is a normalized event probability distribution. Preliminary STAR data reveal concave $R_{\Psi_2}(\Delta S)$ distributions in 200 GeV Au+Au collisions. 
Studies with a multiphase transport (AMPT) and anomalous-viscous Fluid Dynamics (AVFD) models show concave $R_{\Psi_2}(\Delta S)$ distributions for CME signals and convex ones for typical resonance backgrounds. A recent hydrodynamic study, however, indicates concave shapes for backgrounds as well. To better understand these results, we report a systematic study of the elliptic flow ($v_{2}$) and transverse momentum ($p_{T}$) dependences of resonance backgrounds with toy-model simulations and central limit theorem (CLT) calculations. 
It is found that the concavity or convexity of $R_{\Psi_2}(\Delta S)$ depends sensitively on the resonance $v_2$ (which yields different numbers of decay $\pi^+\pi^-$ pairs in the in-plane and out-of-plane directions) and $p_T$ (which affects the opening angle of the decay $\pi^+\pi^-$ pair).
Qualitatively, low $p_{T}$ resonances decay into large opening-angle pairs and result in more ``back-to-back'' pairs out-of-plane, mimicking a CME signal, or a concave $R_{\Psi_2}(\Delta S)$. 
Supplemental studies of $R_{\Psi_3}(\Delta S)$ in terms of the triangular flow ($v_3$), where only backgrounds exist but any CME would average to zero, are also presented.

\end{abstract}


\maketitle

\section{Introduction}

Nontrivial topological gluon fields can form in quantum chromodynamics (QCD) from vacuum fluctuations~\cite{Lee:1974ma}. 
 Interactions with those gluon fields can change the chirality of quarks in local domains where the approximate chiral symmetry is restored~\cite{Lee:1974ma,Morley:1983wr,Kharzeev:1998kz,Kharzeev:2007jp}. 
Quarks of the same chirality in a local domain immersed in a strong magnetic field will move in opposite directions along the magnetic field if they bear opposite charges. This charge separation phenomenon is called the chiral magnetic effect (CME)~\cite{Kharzeev:2007jp,Fukushima:2008xe}.



Heavy-ion collisons provide a suitable environment for the CME to occur: the relativistic spectator protons can create an intense, transient magnetic field~\cite{Kharzeev:2004ey,Bzdak:2011yy,Deng:2012pc,Bloczynski:2012en} roughly perpendicular to the reaction plane (spanned by the impact parameter and beam directions); high energy density can be created in the collision zone and the approximate chiral symmetry may be restored~\cite{Arsene:2004fa,Back:2004je,Adams:2005dq,Adcox:2004mh,Muller:2012zq}; and topological gluon fields can emerge from the QCD vacuum~\cite{Lee:1974ma}. 
Because the observation of the CME will simultaneously support the above pictures,
the detection of such charge separations in heavy-ion collisions is of critical importance.


The common variable that has been used to search for the CME-induced charge separation is the so-called $\Delta\gamma$ variable~\cite{Voloshin:2004vk}. Positive charge-dependent signals have been observed in heavy-ion collisions, qualitatively consistent with the CME~\cite{Abelev:2009ac,Abelev:2009ad,Adamczyk:2014mzf,Adamczyk:2013hsi,Abelev:2012pa}.
However, the $\Delta\gamma$ variable is strongly contaminated by elliptic flow induced correlation backgrounds~\cite{Wang:2009kd,Bzdak:2009fc,Schlichting:2010qia,Zhao:2018ixy, Zhao:2018skm}. In fact, $\Delta\gamma$ measurements in small systems of p+Pb collisions at the CERN Large Hadron Collider (LHC)~\cite{Khachatryan:2016got} and d+Au collisions at the BNL Relativistic Heavy Ion Collider (RHIC)~\cite{Zhao:2017wck,Zhao:2017ckp}, where only backgrounds are expected, reveal large signals comparable to those measured in heavy-ion collisions. With suppression of backgrounds by event-by-event and event-shape-engineering techniques, experimental data~\cite{Adamczyk:2013kcb,Sirunyan:2017quh,Acharya:2017fau} show significantly reduced, consistent-with-zero signals for the CME.

Another variable that has been proposed to detect charge separation is 
the $R_{\Psi_2}(\Delta S)$ variable~\cite{Ajitanand:2010rc, Magdy:2017yjev2}. We call it the \textit{sine} observable. It is defined as follows. 
In each event, let
\begin{equation}\label{Psi2}
\varphi = \phi - \Psi_2
,
\end{equation}
\begin{equation}
\begin{split}
\left\langle S_p \right\rangle = \frac{1}{N_p}\sum_1^{N_p}\sin(\varphi_+),\quad
\left\langle S_n \right\rangle = \frac{1}{N_n}\sum_1^{N_n}\sin(\varphi_-),
\end{split}
\end{equation}
\begin{equation}
\begin{split}
\Delta S_{sep} = \left\langle S_p \right\rangle - \left\langle S_n \right\rangle
,
\end{split}
\end{equation}
where $\phi$ is the particle azimuthal angle in the laboratory frame and $\varphi$ is therefore the azimuthal angle relative to the second-order harmonic plane $\Psi_2$ (as a proxy for the unmeasured reaction plane). Subscripts ($+,-$) indicate the charge sign, and $N_p,N_n$ are the number of particles with positive and negative charge, respectively.
A parallel set of variables is constructed by randomizing the charges of all particles in the event, 
respecting the relative multiplicities of positive and negative particles. Then, according to the randomized charges,
\begin{equation}
\begin{split}
\left\langle S_{p}' \right\rangle = \frac{1}{N_p'}\sum_1^{N_p'}\sin(\varphi'_+),\quad
\left\langle S_{n}' \right\rangle = \frac{1}{N_n'}\sum_1^{N_n'}\sin(\varphi'_-),
\end{split}
\end{equation}
\begin{equation}
\begin{split}
\Delta S_{mix} = \left\langle S_{p}' \right\rangle - \left\langle S_{n}' \right\rangle
,
\end{split}
\end{equation}
where the primes denote quantities for this so-called shuffled event. 
The ratio is formed from the event probability distributions of real events in $\Delta S_{sep}$ and shuffled events in $\Delta S_{mix}$,
\begin{equation}
C_{\Psi_2}(\Delta S) = \frac{N(\Delta S_{sep})}{N(\Delta S_{mix})}
.
\end{equation}

For events with CME signals, charge separation along the magnetic field gives $|\sin\varphi_\pm| \approx 1$ and a maximal difference $\sin\varphi_+-\sin\varphi_- \approx \pm 2$.
The distribution of $N(\Delta S_{sep})$ would therefore become wider than its reference distribution. 
Here, the shuffled event $N(\Delta S_{mix})$ serves as the reference distribution. 
The ratio $C_{\Psi_2}$ is therefore concave for CME~\cite{Ajitanand:2010rc, Magdy:2017yjev2}. 


There can be background sources that change the shape of $C_{\Psi_2}(\Delta S)$. In order to eliminate reaction-plane (RP) independent backgrounds, an analogous variable $C_{\Psi_2}^{\perp}$ is constructed in a way identical to $C_{\Psi_2}$ except changing each $\varphi$ into $\varphi-\pi/2$.
The $R_{\Psi_2}$ variable is defined to be the ratio of $C_{\Psi_2}$ to $C_{\Psi_2}^{\perp}$,
\begin{equation}
R_{\Psi_2}(\Delta S) = \frac{C_{\Psi_2}(\Delta S)}{C_{\Psi_2}^{\perp}(\Delta S)}
.
\end{equation}
The RP-independent backgrounds would cancel in $R_{\Psi_2}(\Delta S)$. Since the CME signal does not affect $C_{\Psi_2}^{\perp}$ significantly because $\sin(\varphi_\pm-\pi/2) \approx 0$, the CME in $C_{\Psi_2}$ would survive in $R_{\Psi_2}(\Delta S)$, making it concave. The RP-dependent backgrounds, such as resonance decays with finite $v_2$, can still affect $R_{\Psi_2}(\Delta S)$. However, they were shown to make $R_{\Psi_2}(\Delta S)$ convex~\cite{Ajitanand:2010rc,Magdy:2017yjev2}.

Preliminary STAR data reveal concave $R_{\Psi_2}(\Delta S)$ distributions in 200 GeV Au+Au collisions~\cite{Roy:2017rs}.
Previous studies using a multiphase transport (AMPT) model where resonance decay background is present but no CME, suggest that $R_{\Psi_2}(\Delta S)$ is convex~\cite{Magdy:2017yjev2}. 
The anomalous-viscous Fluid Dynamics (AVFD) model shows concave $R_{\Psi_2}(\Delta S)$ distributions for CME signals and convex ones for typical resonance backgrounds~\cite{Magdy:2017yjev2}. 
A recent hydrodynamic study, however, indicates concave shapes for backgrounds as well~\cite{Bozek:2017hi}.

To better understand these results, we present a systematic study of resonance backgrounds as functions of the resonance elliptic flow ($v_{2}$) and transverse momentum ($p_{T}$) with toy-model simulations and central limit theorem (CLT) calculations. 
It is found that the concavity or convexity of $R_{\Psi_2}(\Delta S)$ depends sensitively on the resonance $v_2$ (which yields different numbers of decay $\pi^+\pi^-$ pairs in the in-plane and out-of-plane directions) and $p_T$ (which affects the opening angle of the decay $\pi^+\pi^-$ pair).

Supplemental studies in terms of the triangular flow ($v_3$), where only backgrounds exist but any CME would average to zero, are also presented.

\section{Toy-model simulation of resonance backgrounds}


We use a toy model of $\rho$ meson decays to study the behavior of $R_{\Psi_2}(\Delta S)$ as functions of the $\rho$ kinematic variables. The toy model has been used for CME background studies in Ref.~\cite{Wang:2016iov}. It generates events to be composed of primordial pions and $\rho$-decay pions. Their input $p_T$ distributions and $v_2(p_T)$ are obtained from data measurements \cite{Adams:2003cc, Adler:2003qi, Adams:2003xp, Abelev:2008ab, Adams:2004bi, Adare:2010sp, Dong:2004ve, Adamczyk:2015lme, Agashe:2014kda, Abelev:2009gu, Wang:2016iov}. For simplicity, we use the input harmonic plane $\Psi_2$ (as well as $\Psi_3$ discussed in Sec.~\ref{v3simulation}) in our analysis.

\begin{figure*}[th] 
\subfloat{
  \centering
  \includegraphics[width=0.33\linewidth]{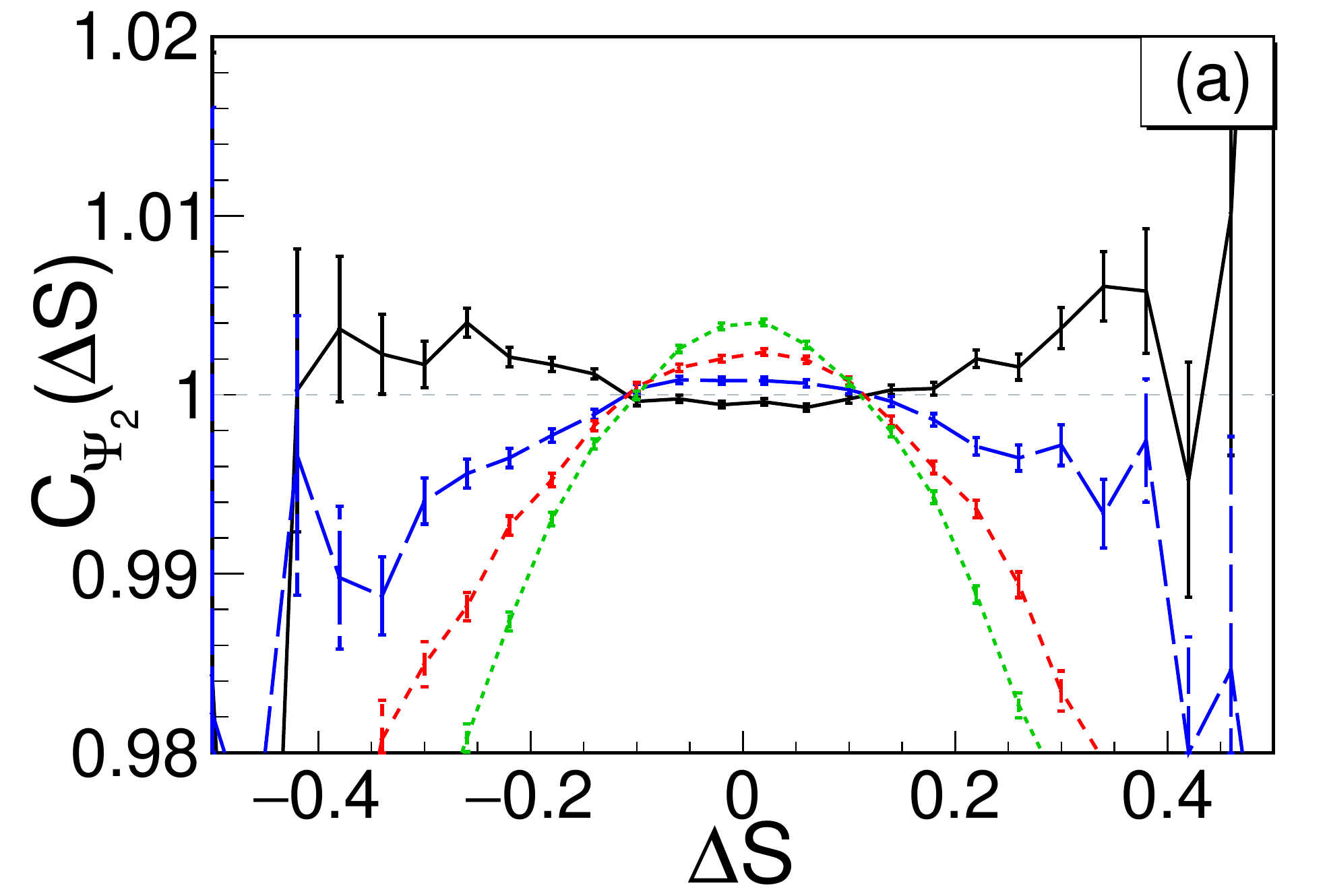}
  \label{Cp_Pi10}
}
\subfloat{
  \centering
  \includegraphics[width=0.33\linewidth]{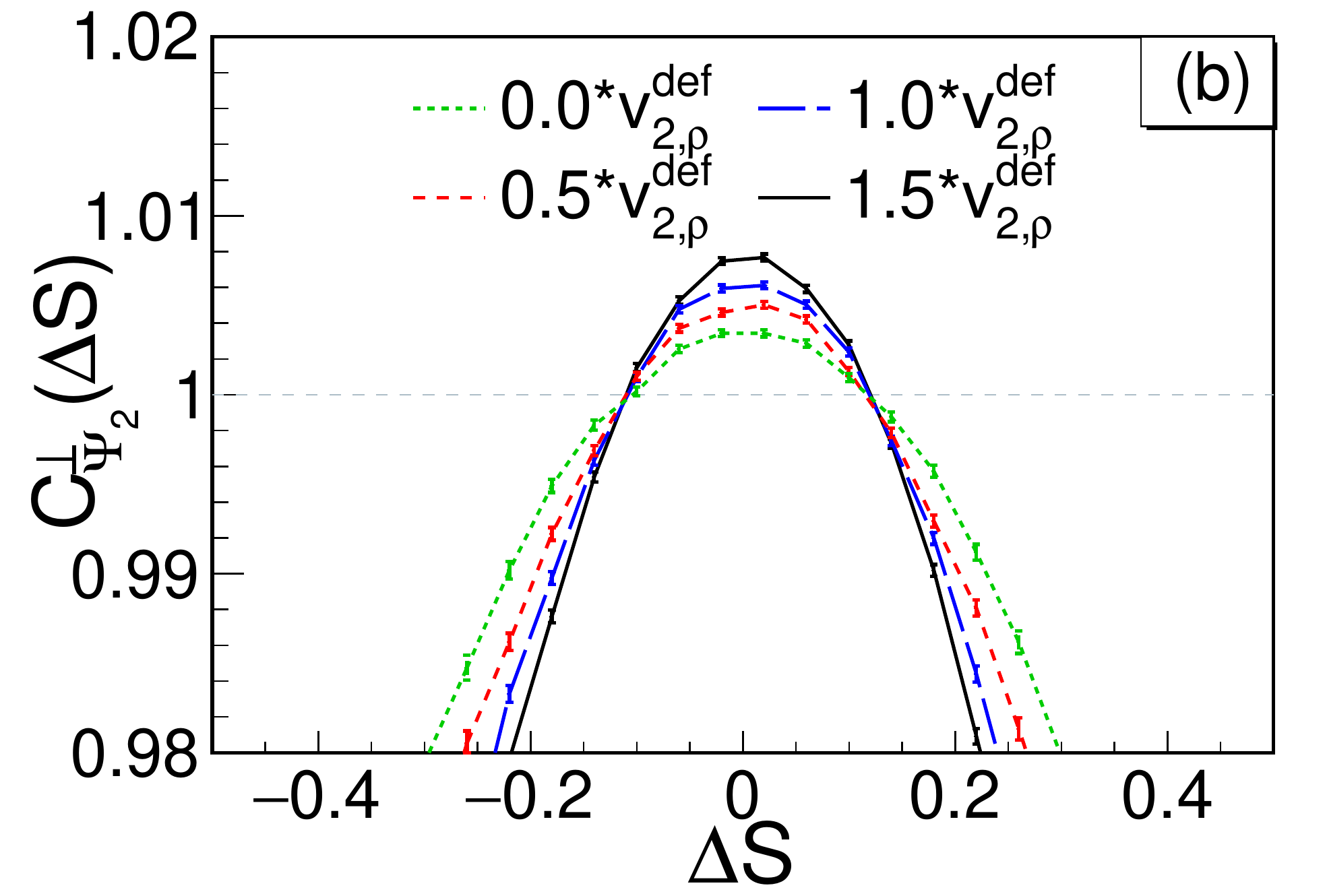}
  \label{Cp_perp_Pi10}
}
\subfloat{
  \centering
  \includegraphics[width=0.33\linewidth]{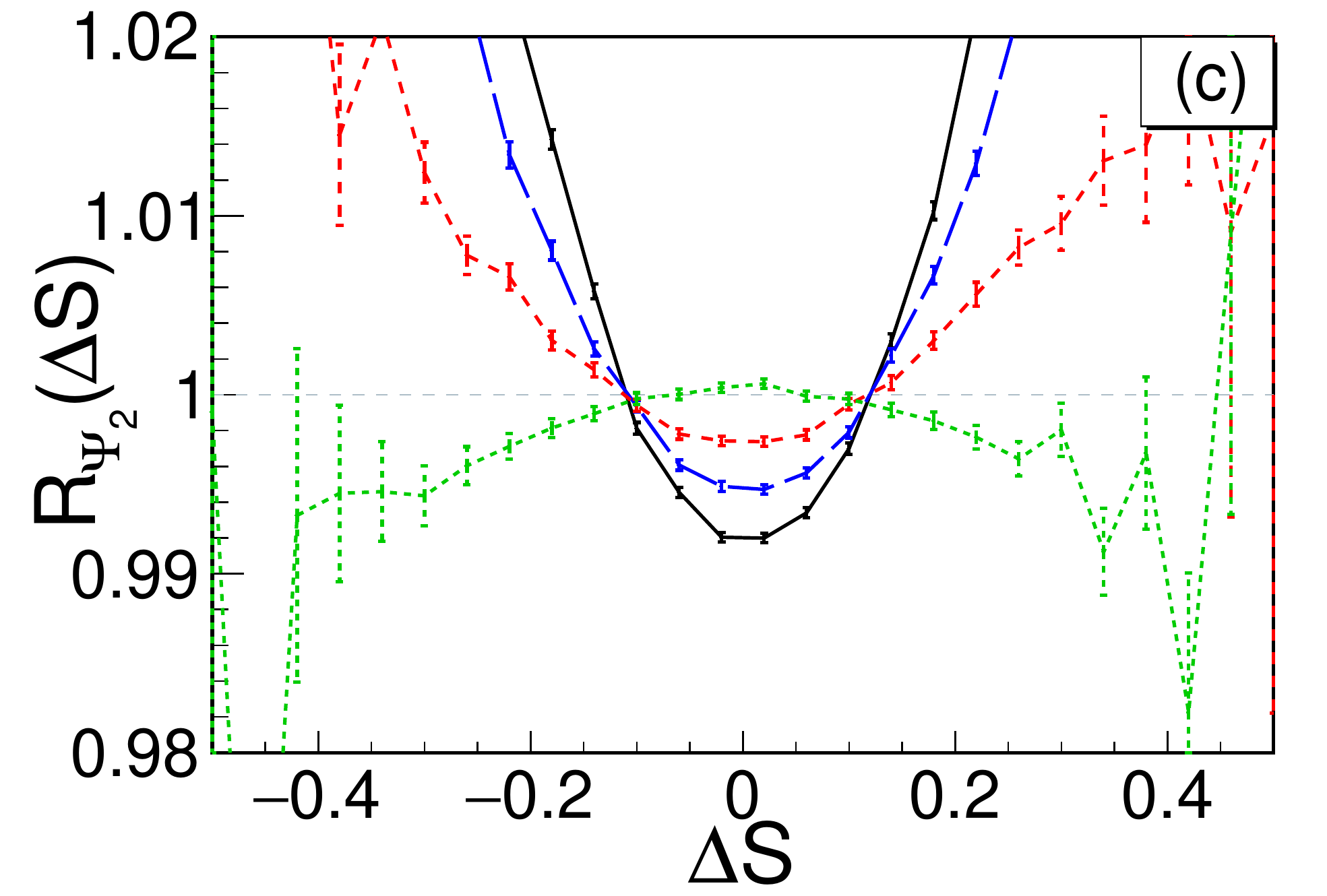}
  \label{R_Pi10}
}

\caption{(color online) Observable distributions for various values of $v_{2,\rho}$ (with $v_{2,\pi}$ fixed to its default distribution). Here, $v_{2,\rho}^{def}$ is the default distribution of $v_{2,\rho}$ obtained from data \cite{Adams:2003cc, Adler:2003qi, Adams:2003xp, Abelev:2008ab, Adams:2004bi, Adare:2010sp, Dong:2004ve, Adamczyk:2015lme, Agashe:2014kda, Abelev:2009gu, Wang:2016iov}. }
\label{v2RhoScan}
\end{figure*}

In order to study the $v_2$ dependence, we scale $v_{2,\rho}$ ($v_2$ of $\rho$) up or down by a $p_T$-independent factor to investigate how $R_{\Psi_2}(\Delta S)$ responds. 
Figure~\ref{v2RhoScan} shows the results; the curve of $C_{\Psi_2}$ becomes more concave when $v_{2,\rho}$ is increased, and $C_{\Psi_2}^{\perp}$ behaves in the opposite way. Subsequently, $R_{\Psi_2}(\Delta S)$ becomes more concave. This behavior can be qualitatively understood as follows.
At the typical resonance $p_T$ in the simulation, the decay daughters are close to each other in azimuthal angle. The numerator of $C_{\Psi_2}$ has the term:
$\sin\varphi_+-\sin\varphi_- \approx \cos\bar{\varphi}\delta\varphi$, where 
$\bar{\varphi}=(\varphi_++\varphi_-)/2$, $\delta\varphi =\varphi_+-\varphi_-$ are the average and difference of the $\pi^\pm$ azimuths, respectively.
When $v_{2,\rho}$ is large, 
$\bar{\varphi}$ will be relatively close to $0$ or $\pi$, and $|\cos\bar{\varphi}|$ will be relatively big. Hence, the $\Delta S$ in the numerator of $C_{\Psi_2}$ has a wider distribution, and accordingly $C_{\Psi_2}$ becomes more concave (see Fig.~\ref{Cp_Pi10}).
Similarly, the numerator of $C_{\Psi_2}^{\perp}$ has the term:
$\sin(\varphi_+-\pi/2)-\sin(\varphi_--\pi/2) \approx \sin\bar{\varphi}\delta\varphi$. When $v_{2,\rho}$ is large, $|\sin\bar{\varphi}|$ will be relatively small and close to $0$, so the $\Delta S$ in the numerator of $C_{\Psi_2}^{\perp}$ has a narrower distribution, and accordingly $C_{\Psi_2}^{\perp}$ becomes more convex (Fig.~\ref{Cp_perp_Pi10}).
Because of the opposite behaviors of $C_{\Psi_2}$ and $C_{\Psi_2}^{\perp}$, we can easily get the dependence of their ratio $R_{\Psi_2}$ on $v_{2,\rho}$: its concavity increases with increasing $v_{2,\rho}$ (Fig.~\ref{R_Pi10}).

\begin{figure}[h]
  \centering
  \includegraphics[width=0.33\textwidth]{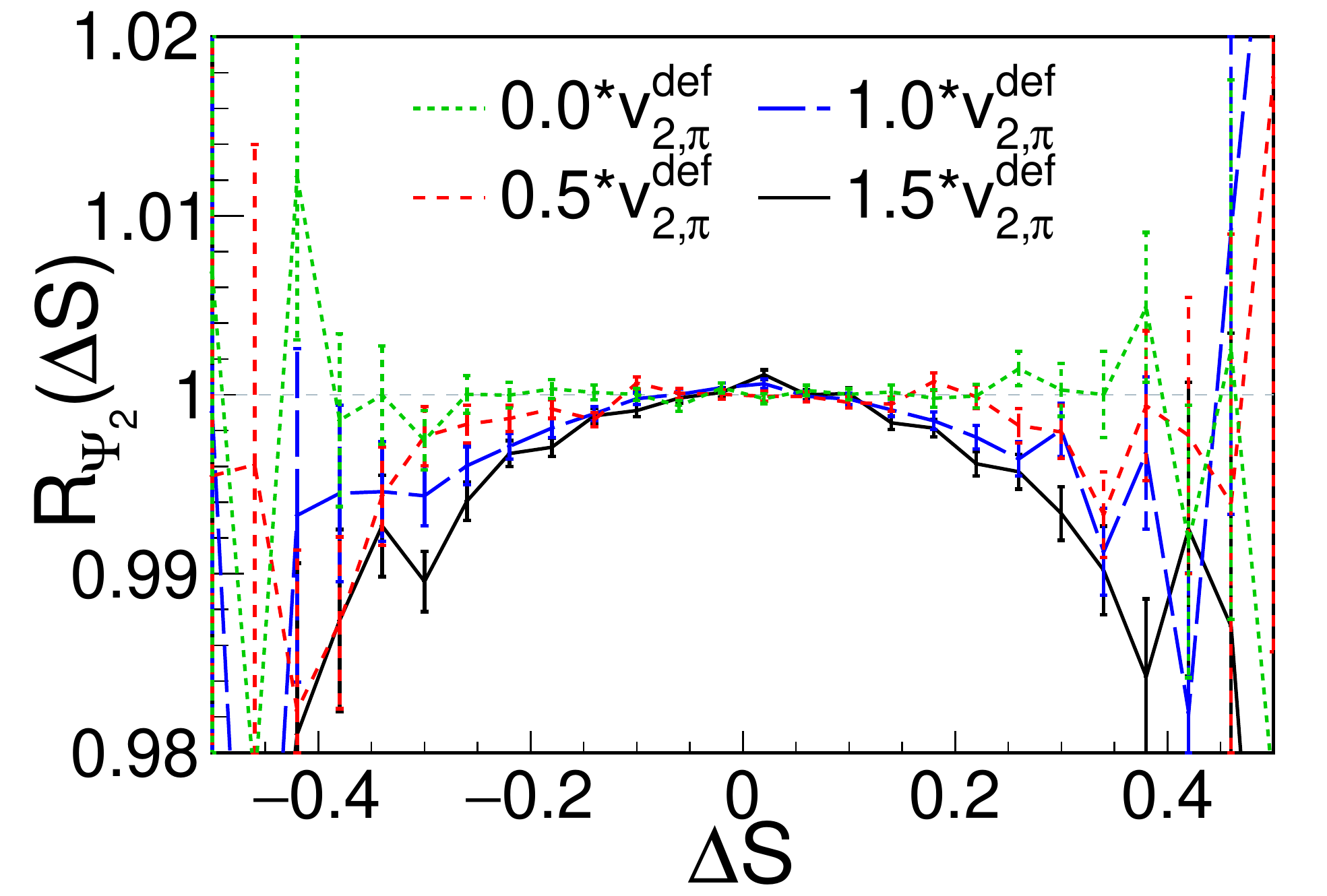}
  \caption{(color online) $R_{\Psi_2}(\Delta S)$ for various values of $v_{2,\pi}$ (with $v_{2,\rho}$ fixed to 0). Here, $v_{2,\pi}^{def}$ is the default distribution of $v_{2,\pi}$ obtained from data \cite{Adams:2003cc, Adler:2003qi, Adams:2003xp, Abelev:2008ab, Adams:2004bi, Adare:2010sp, Dong:2004ve, Adamczyk:2015lme, Agashe:2014kda, Abelev:2009gu, Wang:2016iov}.}
  \label{R_Rho00}
\end{figure}

Note that the curves in Fig.~\ref{R_Pi10} with zero $v_{2,\rho}$ is counterintuitively nonflat. This is due to the finite $v_{2,\pi}$ (primordial pion $v_2$). The $\rho$ decays alter the pion multiplicities which affect $C_{\Psi_2}$ and $C_{\Psi_2}^{\perp}$. The finite $v_{2,\pi}$ breaks the symmetry between $C_{\Psi_2}$ and $C_{\Psi_2}^{\perp}$, resulting in the slightly nonflat $R_{\Psi_2}(\Delta S)$. Figure~\ref{R_Rho00} shows $R_{\Psi_2}$ curves with zero $v_{2,\rho}$ for various values of $v_{2,\pi}$. Only weak dependences on $v_{2,\pi}$ are observed for $R_{\Psi_2}(\Delta S)$ (and also $C_{\Psi_2}$, $C_{\Psi_2}^\perp$). When both $v_{2,\rho}$ and $v_{2,\pi}$ are set to zero, then $R_{\Psi_2}$ is indeed flat. 

\begin{figure*}[th]
\subfloat{
  \centering
  \includegraphics[width=0.33\textwidth]{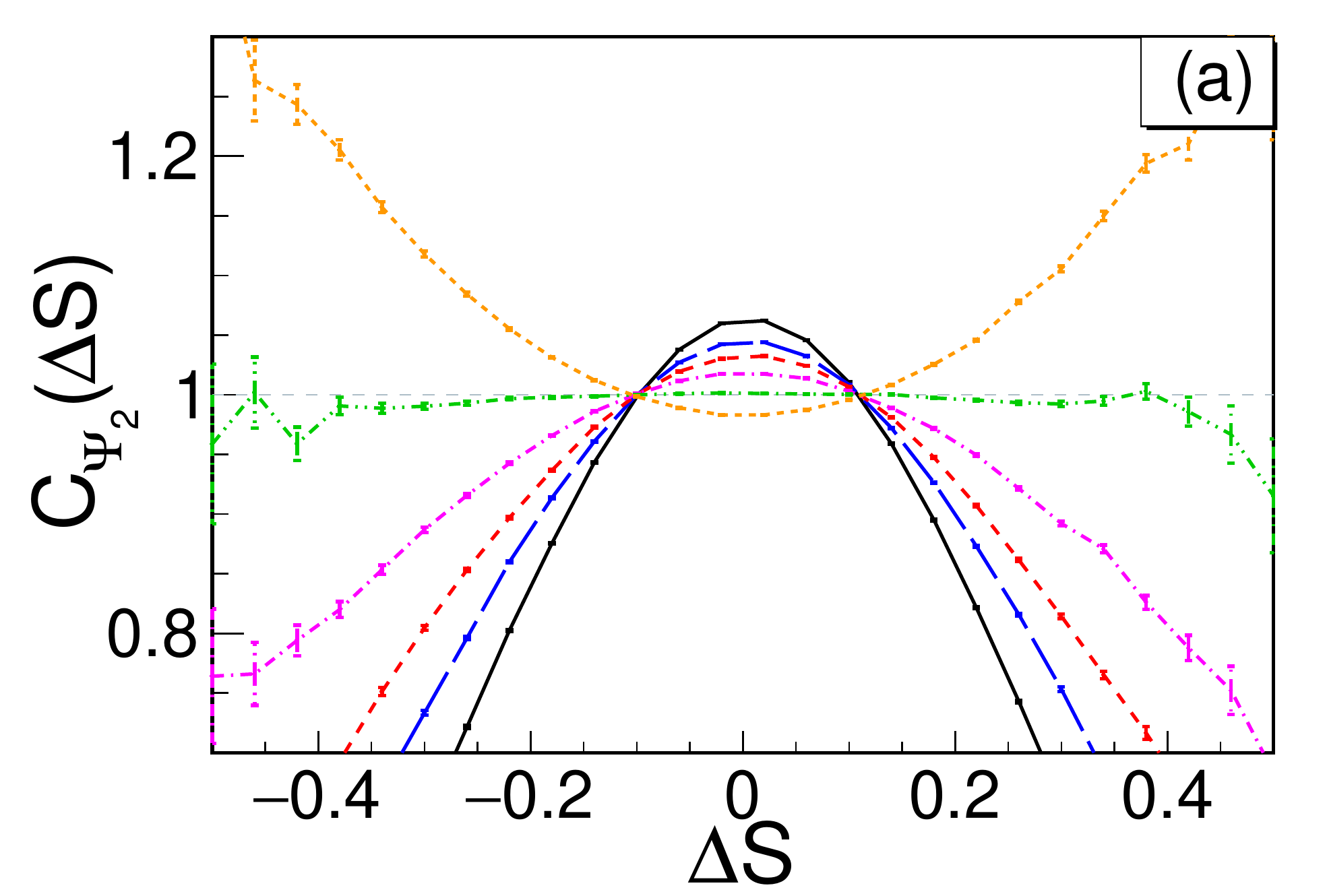}
  \label{Cp_RhoPt}
}
\subfloat{
  \centering
  \includegraphics[width=0.33\textwidth]{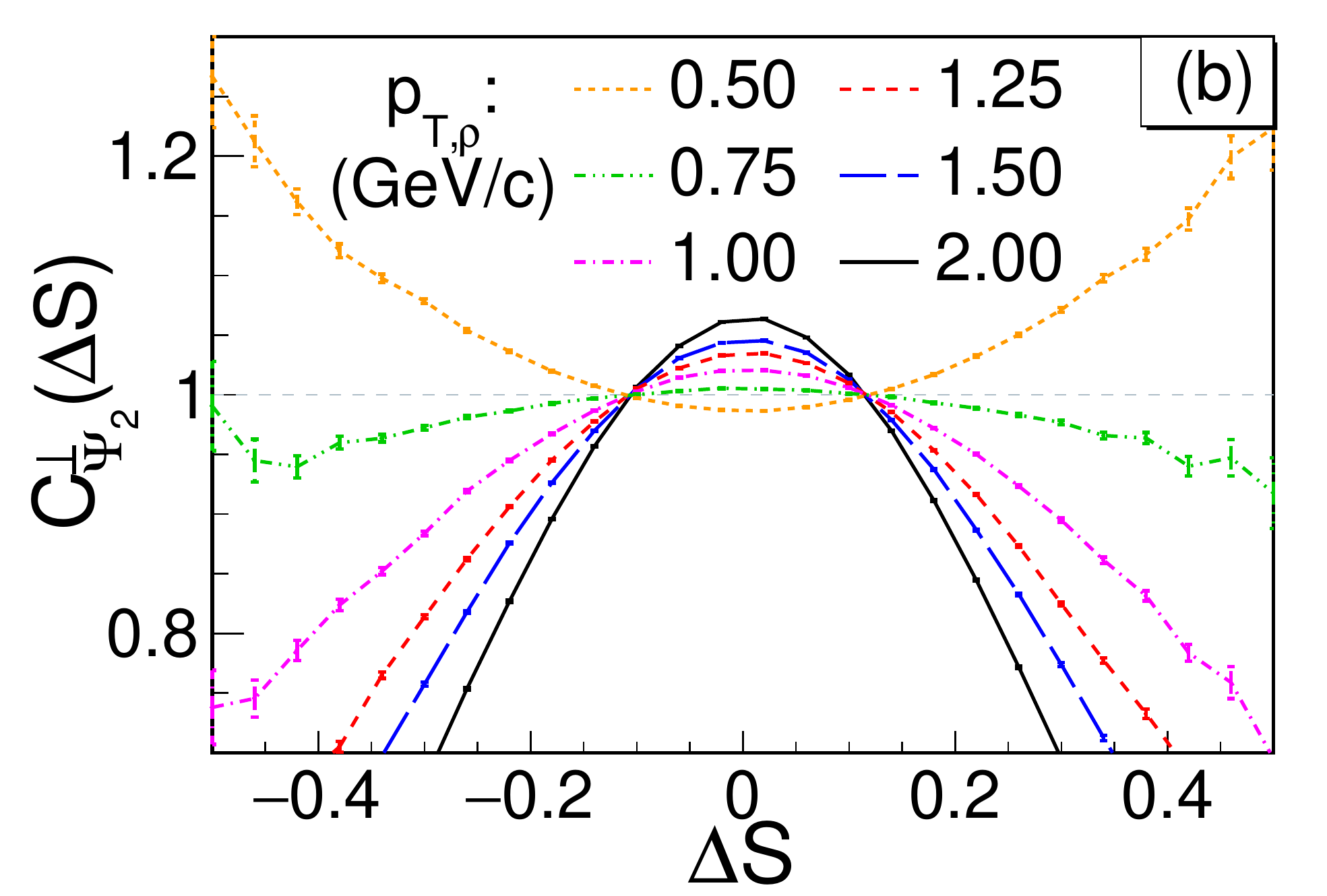}
  \label{Cp_perp_RhoPt}
}
\subfloat{
  \centering
  \includegraphics[width=0.33\textwidth]{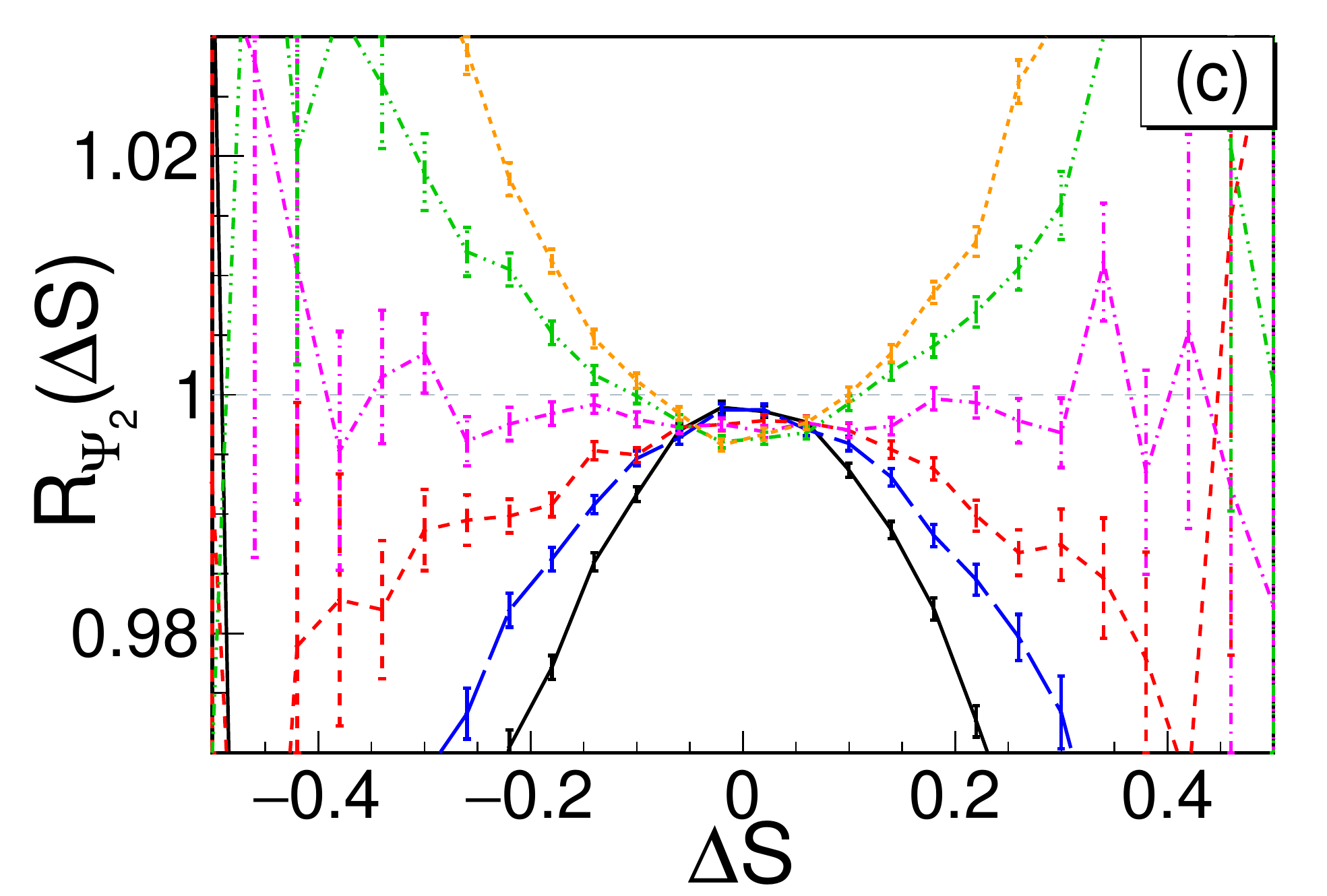}
  \label{R_RhoPt}
}

\caption{(color online) Observable distributions for various values of $p_{T,\rho}$ (with $v_{2,\rho}$ fixed to 0.06).}
\label{v2PtScan}
\end{figure*}

To scan $p_{T,\rho}$ (the $p_T$ of $\rho$), we fix $v_{2,\rho}$ to a specific value 0.06, because otherwise the value of $v_{2,\rho}$ would be affected by the changing $p_{T,\rho}$. 
The $v_{2,\pi}$ and $p_T$ of the primordial pions are given by default.
We find the curves of $C_{\Psi_2}$, $C_{\Psi_2}^{\perp}$, and $R_{\Psi_2}$ to become more convex when $p_{T,\rho}$ increases (Fig.~\ref{v2PtScan}).
This is because of the following. When $p_{T,\rho}$ is large, 
the decay opening angle $\delta\varphi$ is small. 
The $\cos\bar{\varphi}\delta\varphi$ contribution to $\Delta S$ in $C_{\Psi_2}$ and the $\sin\bar{\varphi}\delta\varphi$ contribution to $\Delta S$ in $C_{\Psi_2}^{\perp}$ both become small in magnitude, so the distributions of $\Delta S$ in both $C_{\Psi_2}$ and $C_{\Psi_2}^{\perp}$ become narrower. The reshuffled $\Delta S$ in the denominators of $C_{\Psi_2}$ and $C_{\Psi_2}^{\perp}$ are not as sensitive to the $\delta\varphi$ change as the numerators. Thus, the shapes of $C_{\Psi_2}$ and $C_{\Psi_2}^{\perp}$ both become more convex. Since the change in $\cos\bar{\varphi}\delta\varphi$ is larger than in $\sin\bar{\varphi}\delta\varphi$ with increasing $p_T$ for $\bar{\varphi}$ close to the reaction plane, the narrowing in $C_{\Psi_2}$ is more significant, so $R_{\Psi_2}$ becomes more convex.

Another way to explain the $R_{\Psi_2}$ change is as follows. When $p_{T,\rho}$ is high, the two decay daughters are close to each other and preferentially close to the reaction plane because of the finite $v_{2,\rho}$. This is characteristic of the CME background. At low $p_{T,\rho}$, the two daughters are preferentially more perpendicular to the RP because of the large decay opening angle. This case resembles the CME signal, so the $R_{\Psi_2}$ curves with lower $p_{T,\rho}$ becomes more concave, just like how CME signal would behave. 
For our typical $p_T$ distribution from data, the high $p_{T,\rho}$ case wins over the case with low $p_{T,\rho}$.

\begin{figure*}[h]
\subfloat{
  \centering
  \includegraphics[width=0.33\textwidth]{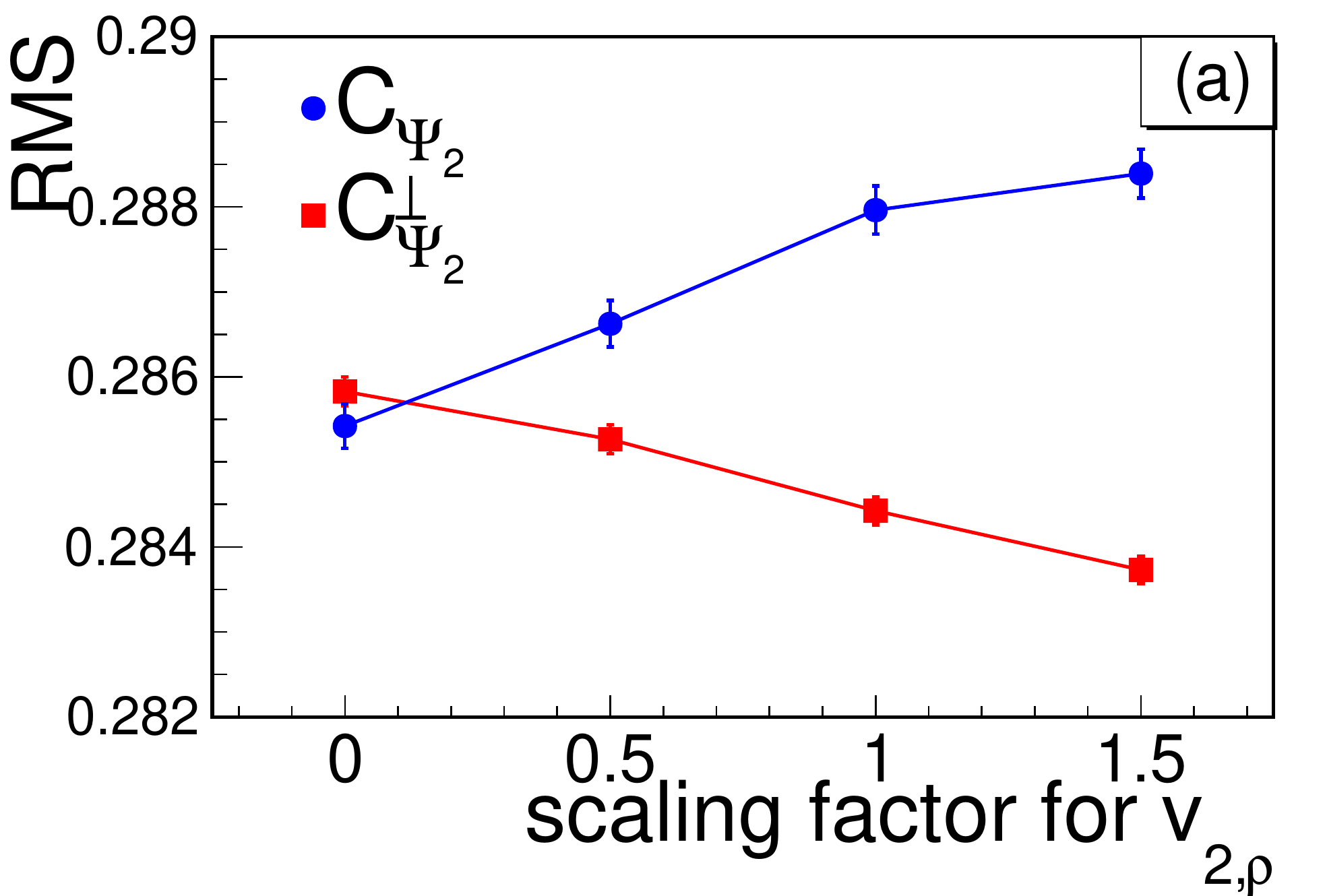}
  \label{RMS_Pi10}
}
\quad\quad
\subfloat{
  \centering
  \includegraphics[width=0.33\textwidth]{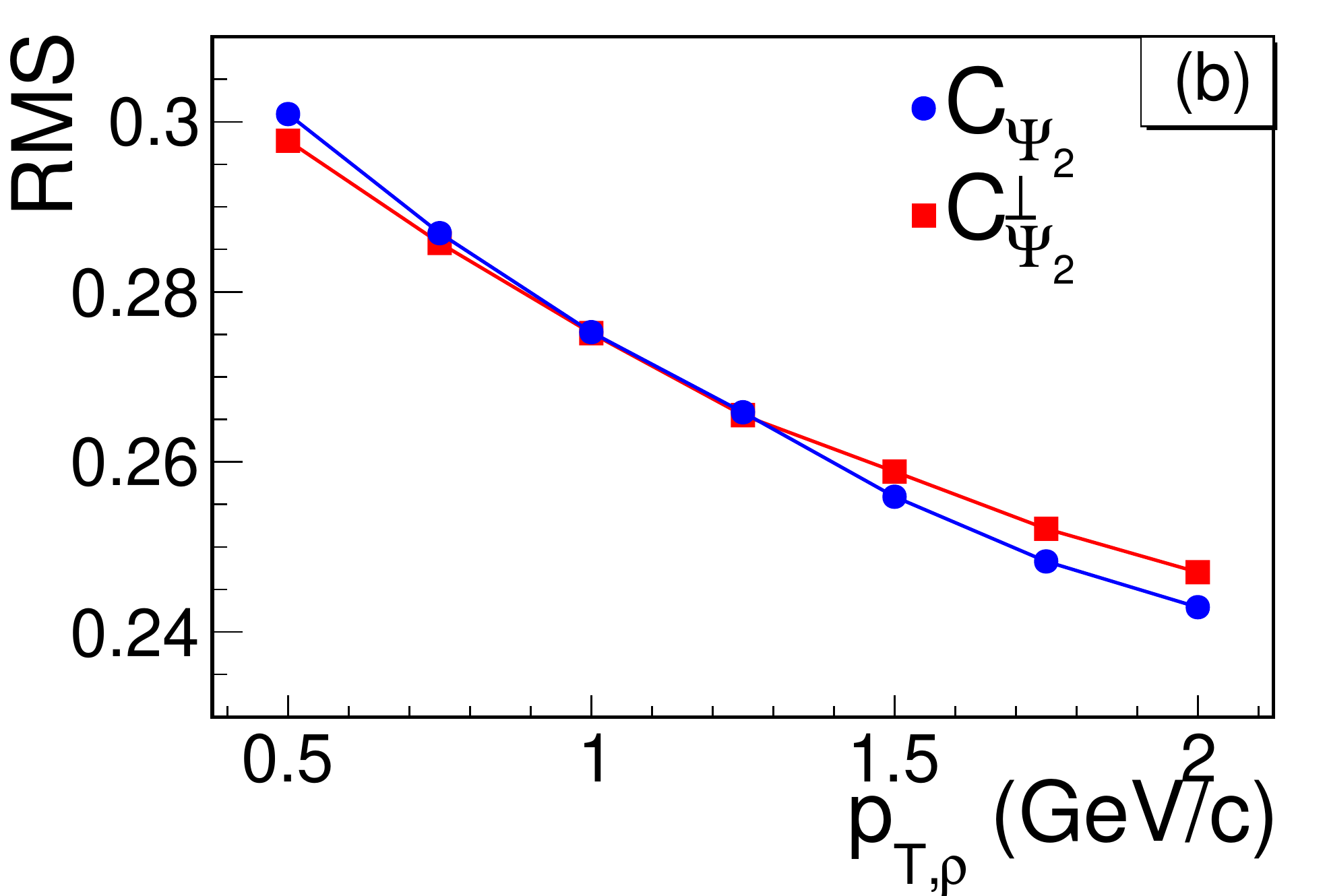}
  \label{RMS_Rho10}
}

\caption{(color online) RMS of $C_{\Psi_2}$ and $C_{\Psi_2}^{\perp}$ depending on $v_{2,\rho}$ and $p_{T,\rho}$. (a) RMS of $C_{\Psi_2}$ and $C_{\Psi_2}^{\perp}$ (shown in Fig.~\ref{v2RhoScan}) depending on $v_{2,\rho}$ (with $v_{2,\pi}$ fixed to its default distribution). (b) RMS of $C_{\Psi_2}$ and $C_{\Psi_2}^{\perp}$ (shown in Fig.~\ref{v2PtScan}) depending on $p_{T,\rho}$ (with $v_{2,\rho}$ fixed to 0.06 and $v_{2,\pi}$ fixed to its default distribution).}
\label{RMS2}
\end{figure*}

The behaviors of $C_{\Psi_2}$ and $C_{\Psi_2}^{\perp}$ are recapitulated in Fig.~\ref{RMS2} by  the RMS (root mean square) of $C_{\Psi_2}$ and $C_{\Psi_2}^{\perp}$.

We summarize our main findings as follows:
\begin{itemize}
\item The curve of $C_{\Psi_2}$ becomes more concave when $v_{2,\rho}$ increases, and $C_{\Psi_2}^{\perp}$ more convex, rendering a more concave $R_{\Psi_2}$.
\item The shapes of the observables ($C_{\Psi_2}$, $C_{\Psi_2}^{\perp}$, and $R_{\Psi_2}$) are only weakly dependent on $v_{2,\pi}$.
\item The curves of $C_{\Psi_2}$ and $C_{\Psi_2}^{\perp}$ become more convex when $p_{T,\rho}$ increases. The effect is more significant in $C_{\Psi_2}$, rendering a more convex $R_{\Psi_2}$.
\end{itemize}


\section{Supplemental studies using $v_3$} \label{v3simulation}

The CME is a charge separation with respect to the RP (or the $v_2$ harmonic plane $\Psi_2$). The CME-induced charge separation must be zero with respect to the third order harmonic plane because of its random orientation relative to $\Psi_2$. Resonance backgrounds, on the other hand, should be still finite with respect to $\Psi_3$. In this section, we verify this with our toy model simulation.


In term of $v_3$, the reference azimuthal angle is the third harmonic plane:
\begin{equation} \label{Psi3}
\varphi=\phi-\Psi_3
.
\end{equation}
There have been two different ways to define the sine observables for $v_3$, and both are similar to the definition of the observables for $v_2$.

\textbf{A. }
For the first definition~\cite{Magdy:2017yjev2}, 
one changes $\Psi_2$ into $\Psi_3$ for $\varphi$ (see Eqs.~\ref{Psi2},~\ref{Psi3}) and replaces $-\pi/2$ by $-\pi/3$ for $\Delta S$ (both $\Delta S_{sep}$ and $\Delta S_{mix}$) in $C_{\Psi_3}^{\perp}$,
\begin{equation}
C_{\Psi_3}^{\perp}: \quad \Delta S = \frac{1}{N_p}\sum_1^{N_p} \sin\left(\varphi_+-\frac{\pi}{3}\right)-\frac{1}{N_n}\sum_1^{N_n} \sin\left(\varphi_--\frac{\pi}{3}\right)
.
\end{equation}

\textbf{B. }
For the second definition~\cite{Bozek:2017hi}, one still changes $\Psi_2$ into $\Psi_3$ for $\varphi$. In addition, one adds a factor $3/2$ in front of the azimuths,
\begin{equation}
\begin{split}
C_{\Psi_3}:& \quad \Delta S = \frac{1}{N_p} \sum_1^{N_p} \sin\left(\frac{3}{2}\varphi_+\right)-\frac{1}{N_n} \sum_1^{N_n} \sin\left(\frac{3}{2}\varphi_-\right)
,
\\
C_{\Psi_3}^{\perp}:& \quad \Delta S = \frac{1}{N_p} \sum_1^{N_p} \sin\left(\frac{3}{2}\varphi_+-\frac{\pi}{2}\right)-\frac{1}{N_n} \sum_1^{N_n} \sin\left(\frac{3}{2}\varphi_--\frac{\pi}{2}\right)
.
\end{split}
\end{equation}

\begin{figure*}[th]
\subfloat{
  \centering
  \includegraphics[width=0.33\textwidth]{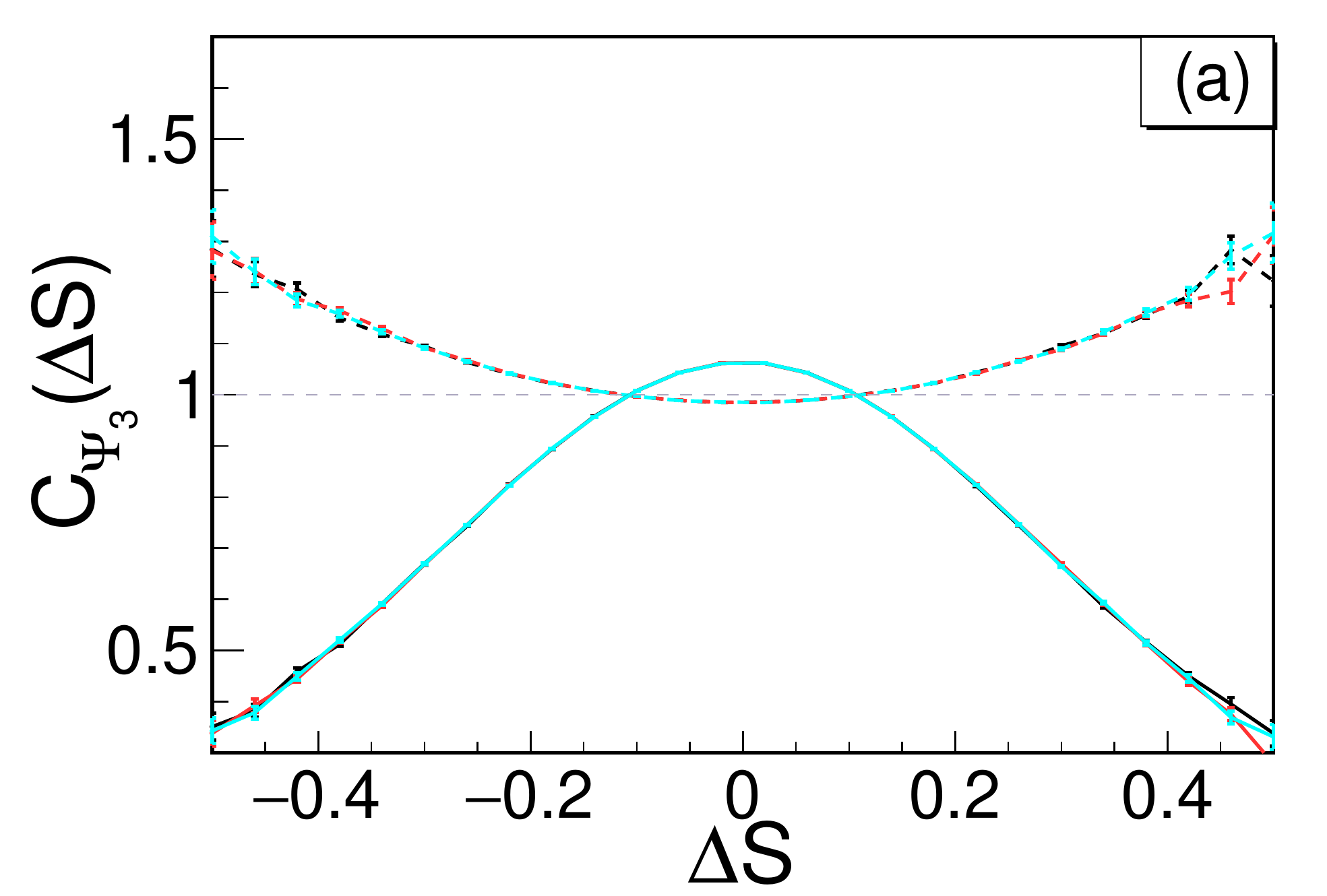}
  \label{N3Cp}
}
\subfloat{
  \centering
  \includegraphics[width=0.33\textwidth]{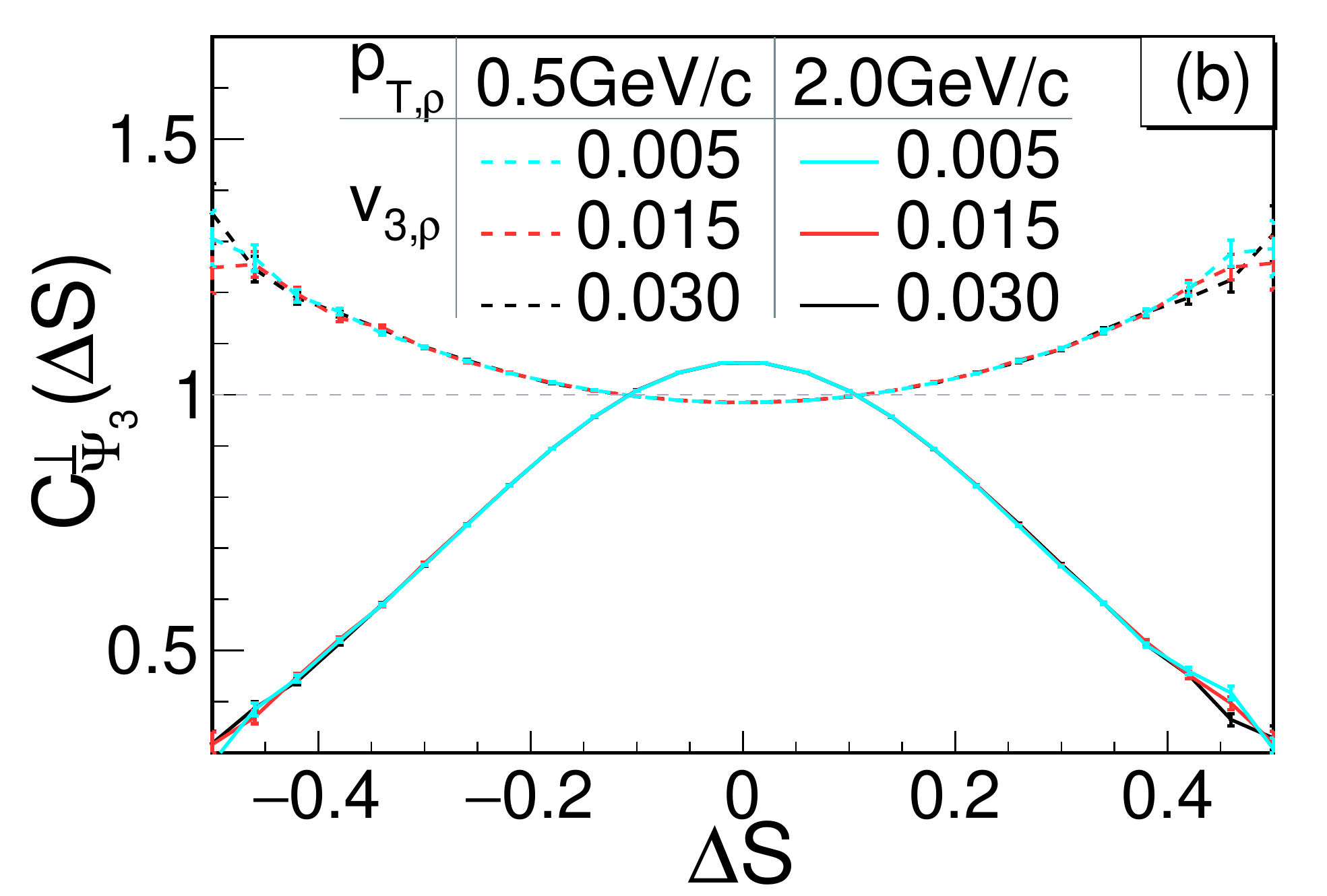}
  \label{N3Cp_perp}
}
\subfloat{
  \centering
  \includegraphics[width=0.33\textwidth]{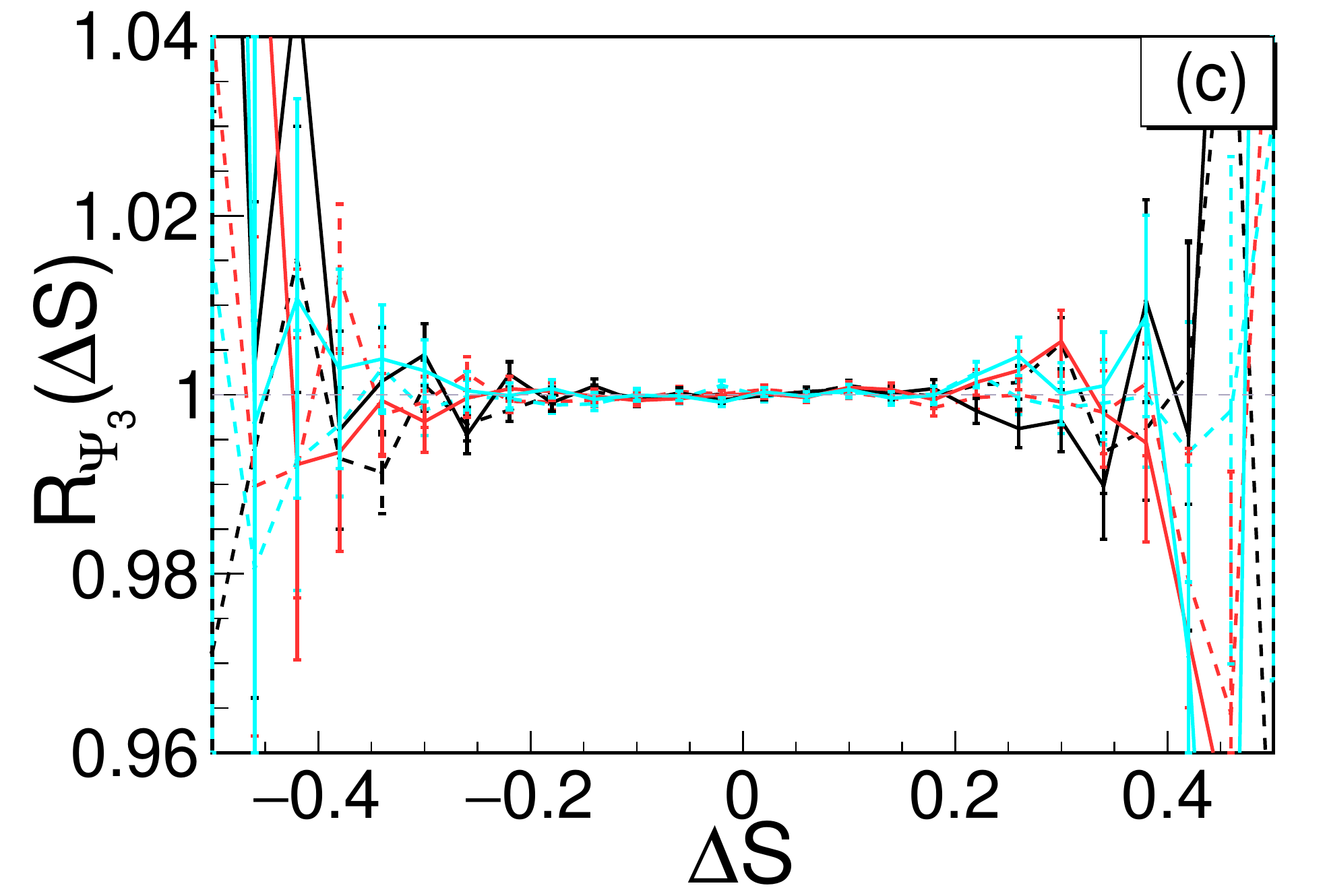}
  \label{N3R}
}

\caption{(color online) Definition \textbf{A}: observable distributions for various values of $v_{3,\rho}$ and $p_{T,\rho}$ (with $v_{3,\pi}$ fixed to $0$). The $C_{\Psi_3}$ and $C_{\Psi_3}^{\perp}$ curves, with the same $p_{T,\rho}$ but various $v_{3,\rho}$, are very close to each other in panels (a) and (b) (concave dashed lines for low $p_{T,\rho}$, and convex solid lines for high $p_{T,\rho}$). }
\label{v3definition1}
\end{figure*}

\begin{figure*}[th]
\subfloat{
  \centering
  \includegraphics[width=0.33\textwidth]{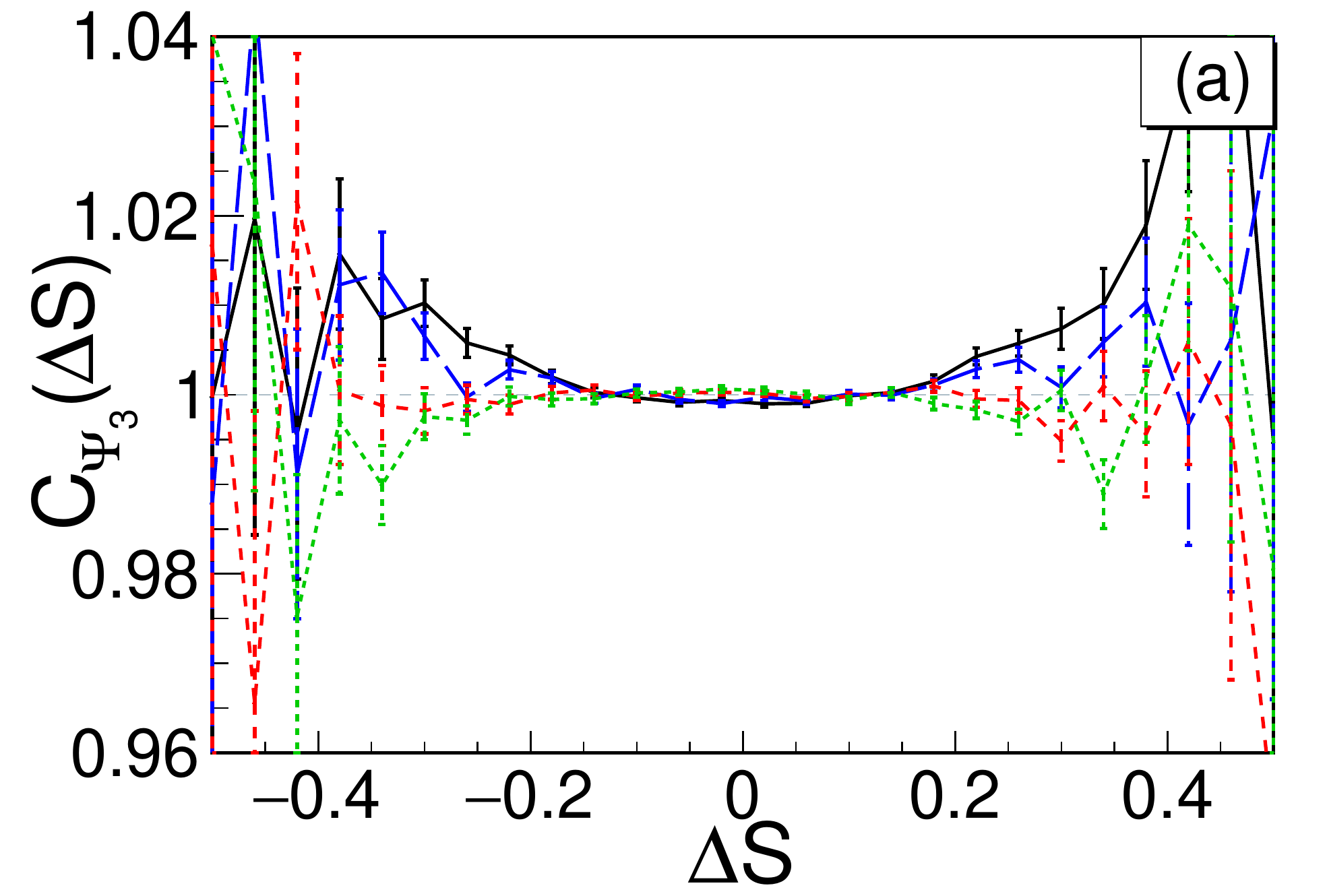}
  \label{B3Cp}
}
\subfloat{
  \centering
  \includegraphics[width=0.33\textwidth]{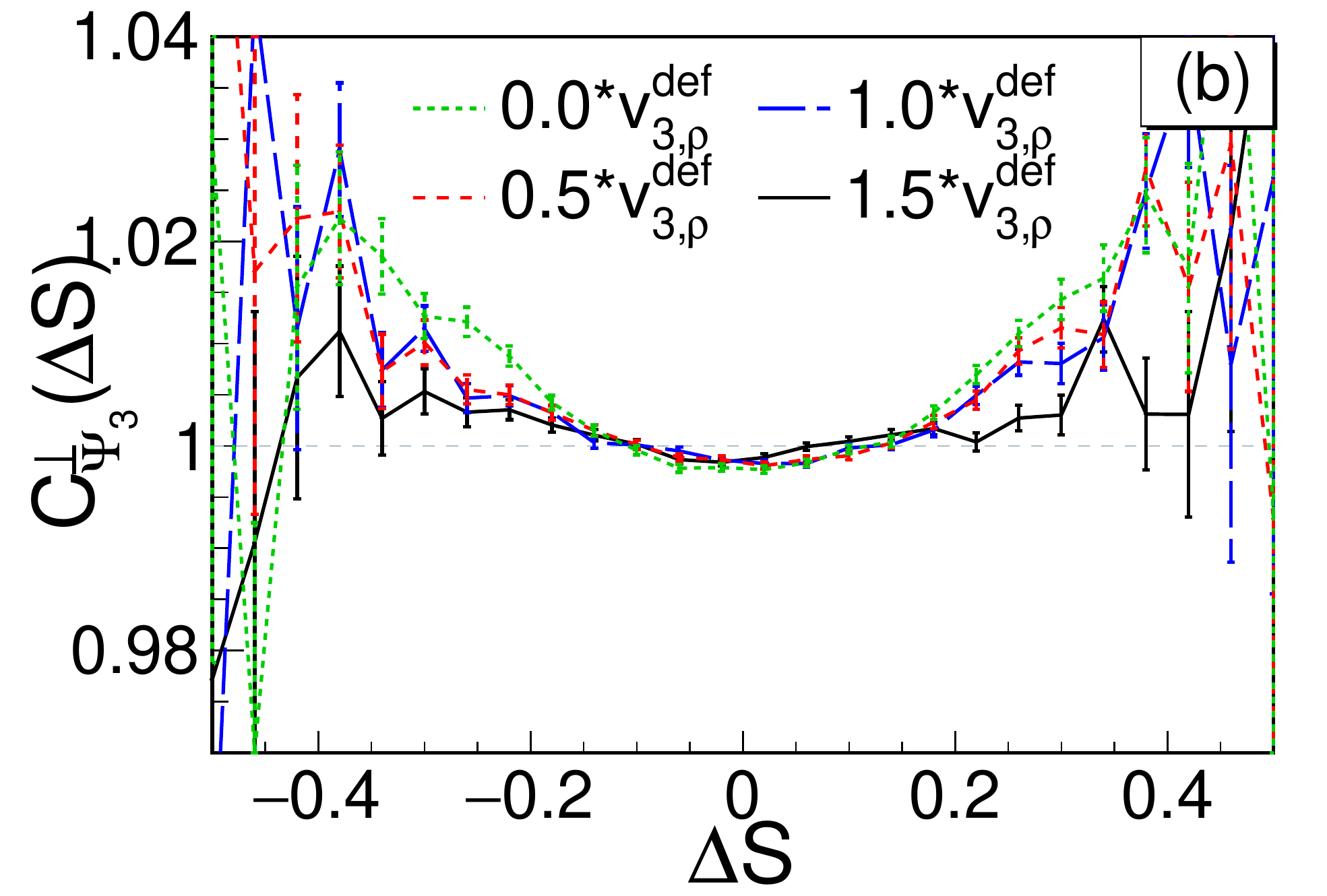}
  \label{B3Cp_perp}
}
\subfloat{
  \centering
  \includegraphics[width=0.33\textwidth]{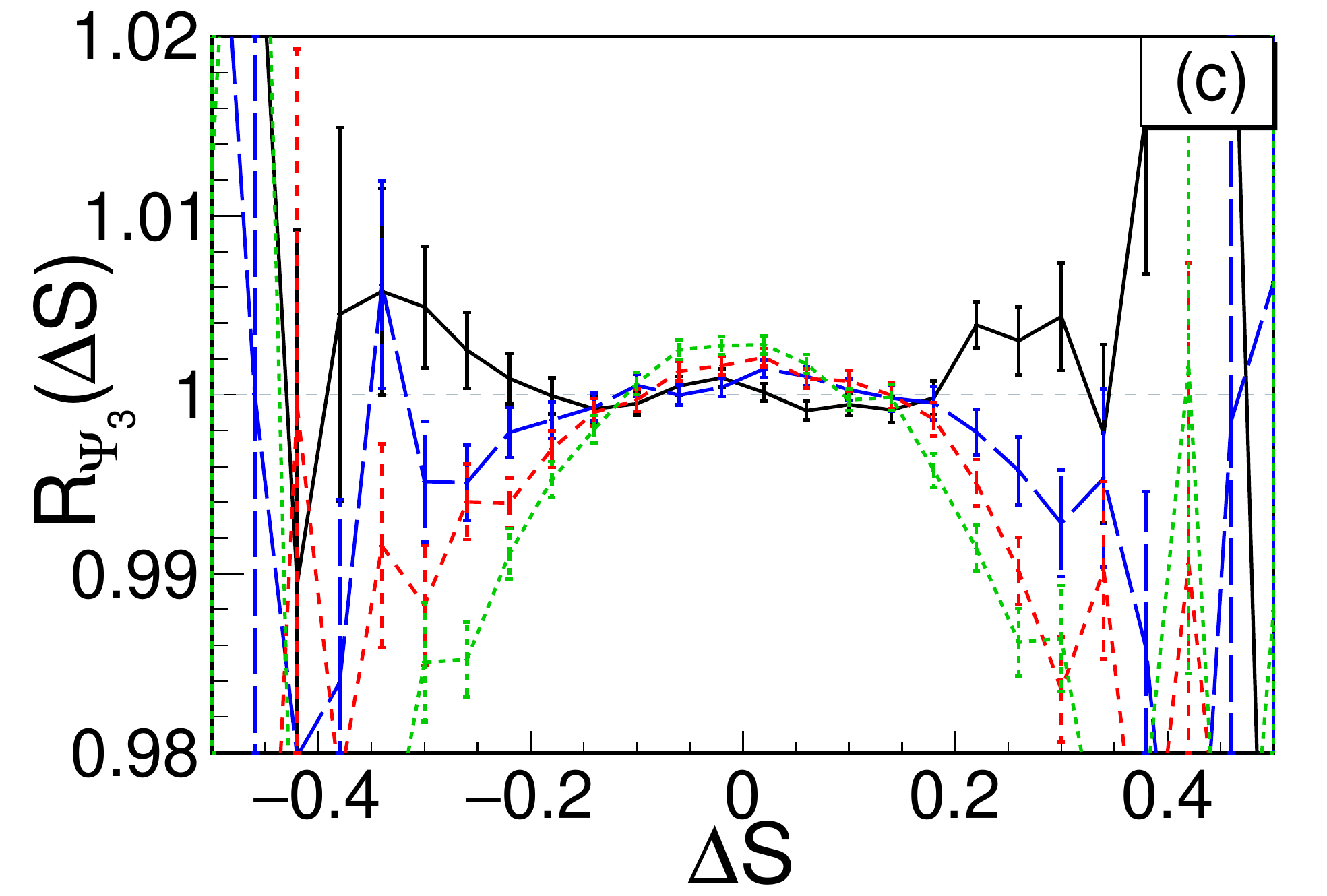}
  \label{B3R}
}

\subfloat{
  \centering
  \includegraphics[width=0.33\textwidth]{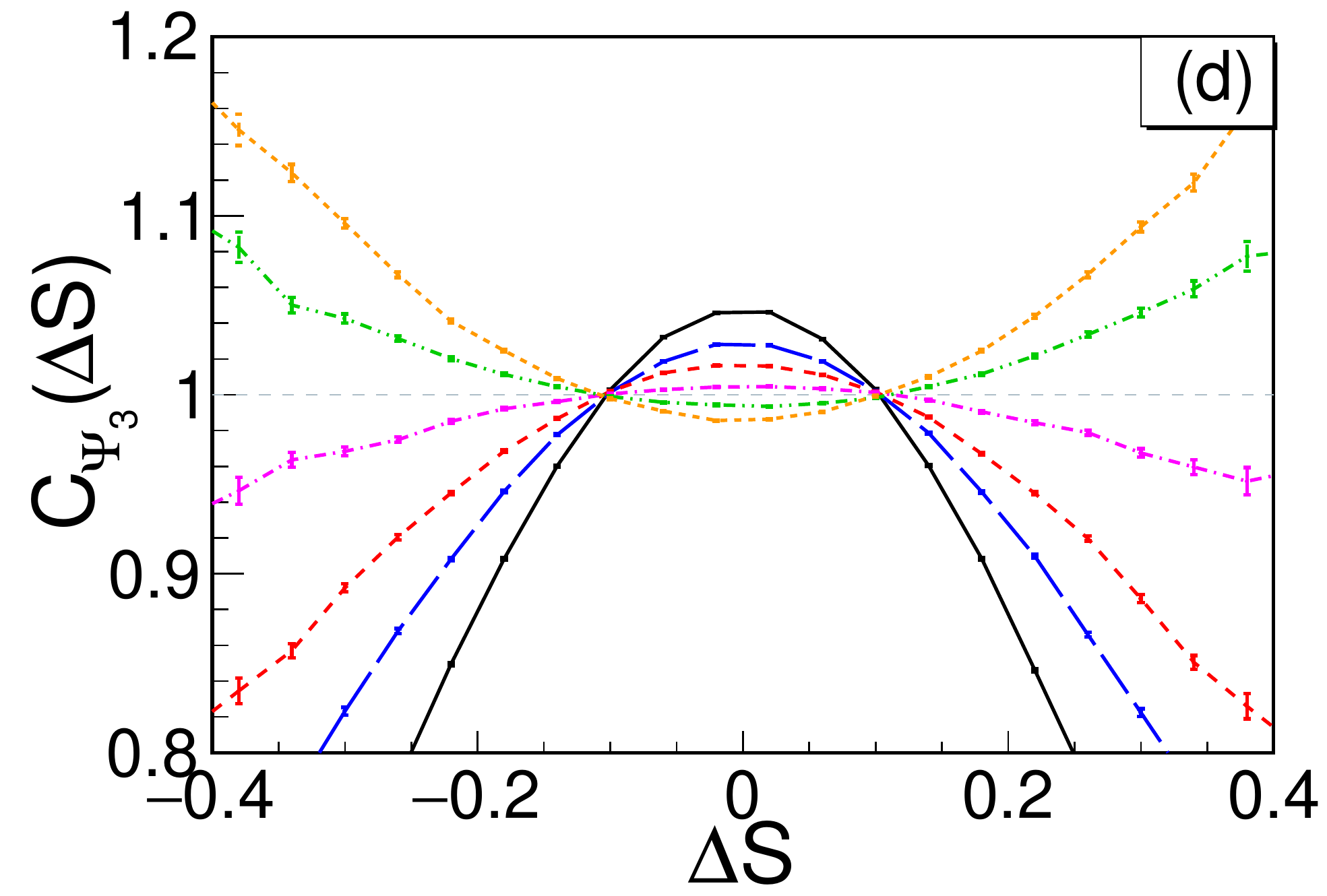}
  \label{B3Cp_pT}
}
\subfloat{
  \centering
  \includegraphics[width=0.33\textwidth]{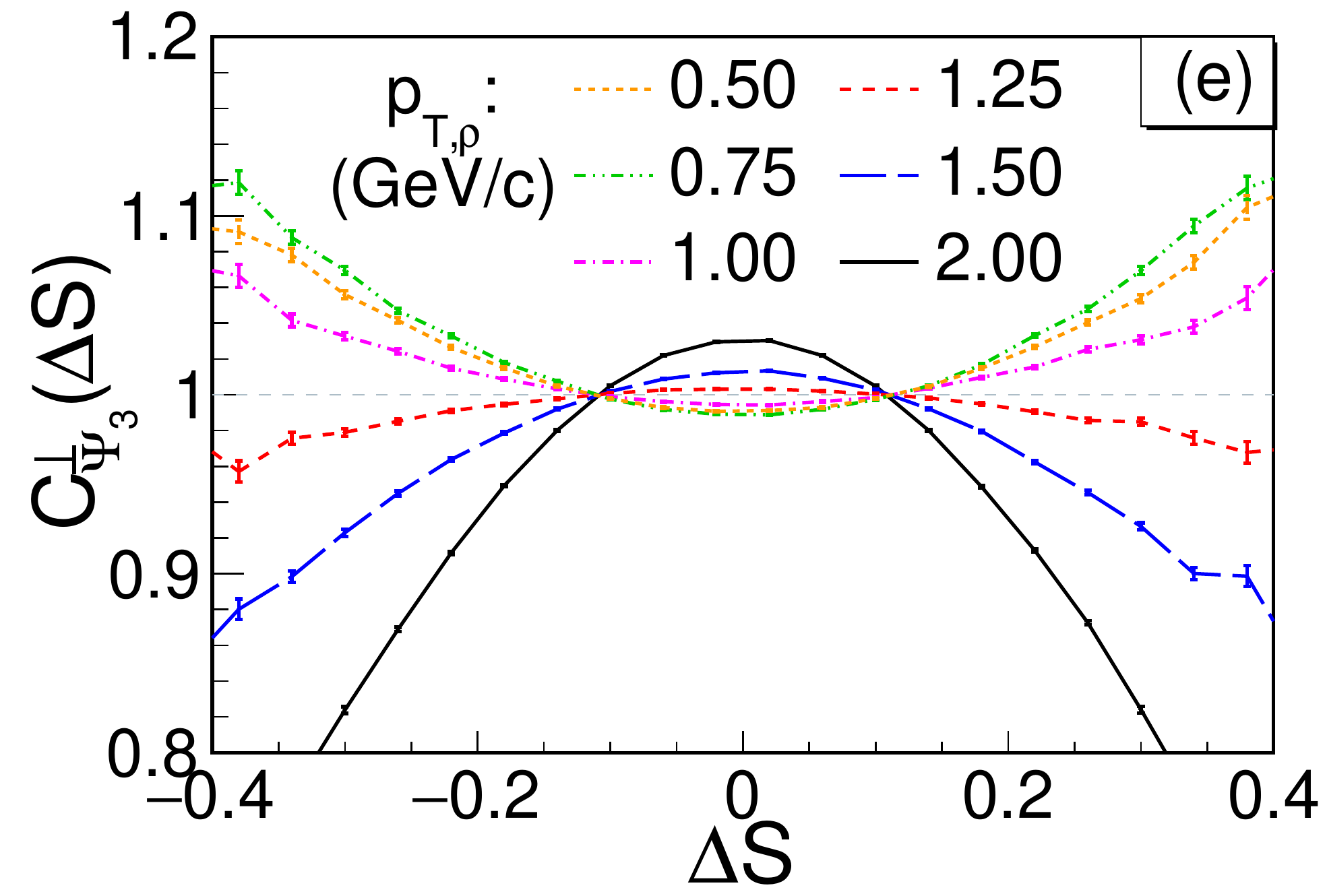}
  \label{B3Cp_perp_pT}
}
\subfloat{
  \centering
  \includegraphics[width=0.33\textwidth]{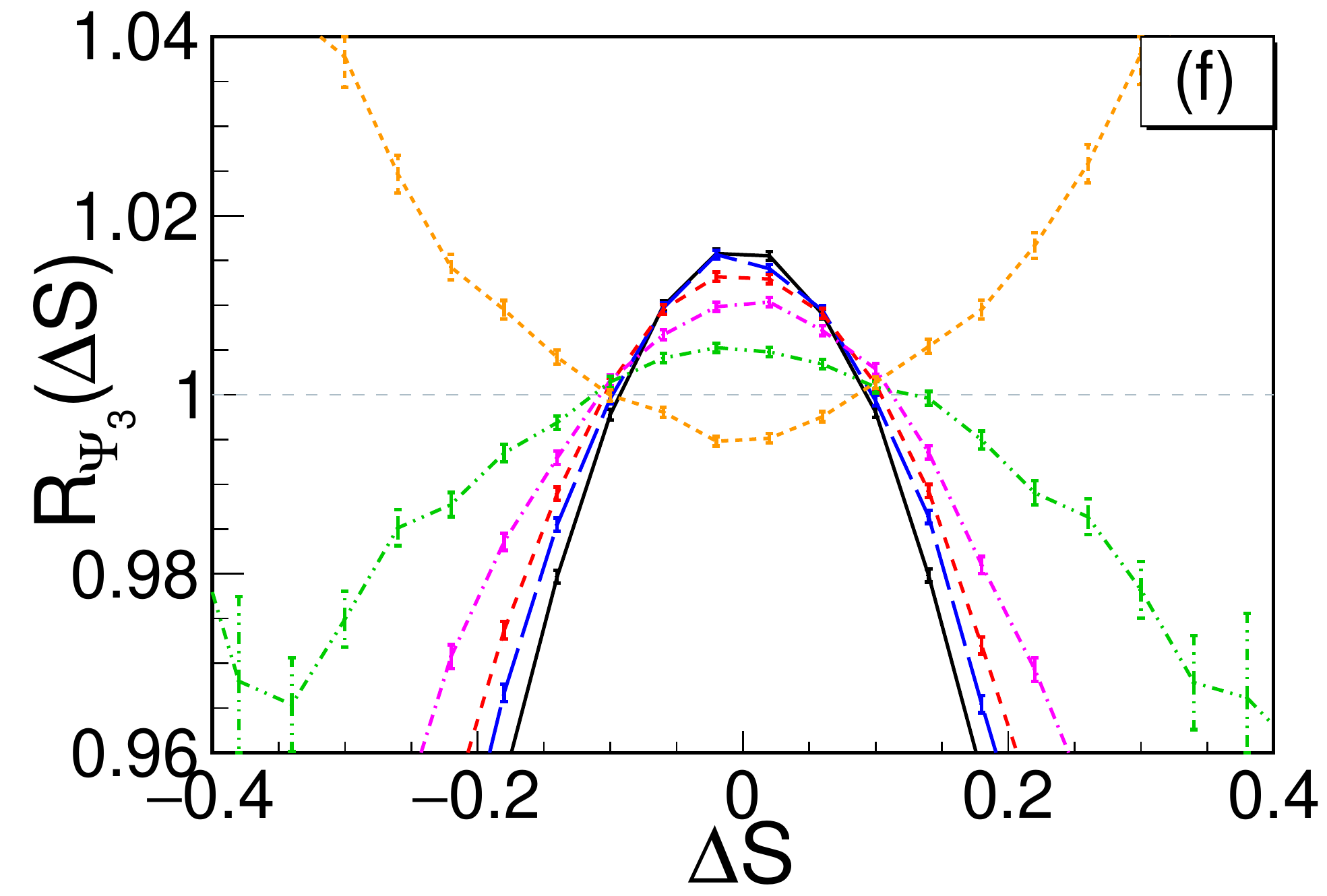}
  \label{B3R_pT}
}
\caption{(color online) Definition \textbf{B}: The upper plots (a)--(c) show observable distributions for various values of $v_{3,\rho}$ (with $v_{3,\pi}$ fixed to $0$). Here, $v_{3,\rho}^{def}$ is the default distribution of $v_{3,\rho}$ approximated by $0.5v_{2,\rho}^{def}$ (Eq.~\ref{default v3Rho definition}).
The lower plots (d)--(f) show observable distributions for various values of $p_{T,\rho}$ (with $v_{3,\rho}$ fixed to 0.03 and $v_{3,\pi}$ fixed to $0$).}
\label{v3 definition 2}
\end{figure*}

We use the toy Monte Carlo simulation to investigate $R_{\Psi_3}(\Delta S)$ of those two definitions. The toy simulation generates primordial $\pi^+, \pi^-$ and $\rho$ with the experimental $p_T$ spectra but with only $v_3$ of the $\rho$. Since $\Psi_2$ and $\Psi_3$ are uncorrelated, including non-zero $v_2$ does not change the results. 
Including a finite $v_3$ for the primordial pions does not have significant effect.
The default function of $v_{3,\rho}(p_T)$ is approximated by that of $v_{2,\rho}(p_T)$ but with half magnitude, i.e.
\begin{equation} \label{default v3Rho definition}
v_{3,\rho}(p_T) = \frac{1}{2}v_{2,\rho}(p_T)
.
\end{equation} 
We constrain the azimuthal range to be $[0,2\pi)$ in our simulation.
As will be discussed later in Sec.~\ref{CLTv3}, the sine observables of Definition~\textbf{B} unfortunately depend on which periodic range is used, suggesting Definition~\textbf{B} is not a physically correct definition.
The simulation results are shown in Figs.~\ref{v3definition1} and \ref{v3 definition 2}.


We make the following observations:
\begin{itemize}
\item In Definition \textbf{A}, $R_{\Psi_3}$ is always flat.

By Definition \textbf{A} itself, $R_{\Psi_3}$ should always be flat, as follows. 
The Probability Density Function (PDF) is $f(\varphi)=(1+2v_3\cos(3\varphi))/(2\pi)$, whose period is $2\pi/3$. In the definition of $C_{\Psi_3}^{\perp}$, $\varphi$ is shifted by $\pi/3$ clockwise, $\Delta S(\varphi) \rightarrow \Delta S(\varphi -\pi/3)$. If we keep shifting $\Delta S$ by another period in the same direction, we would not change the distribution of $\Delta S$ in $C_{\Psi_3}^{\perp}$, which means $\Delta S(\varphi - \pi/3)$ and $\Delta S(\varphi-\pi)$ have the same distribution. From the Definition \textbf{A}, we also know that $\Delta S(\varphi-\pi) = -\Delta S(\varphi)$. Because the distribution of $\Delta S(\varphi)$ is symmetric about $\Delta S = 0$, $\Delta S(\varphi)$ and $-\Delta S(\varphi)$ have the same distribution as well. Thus, $\Delta S(\varphi)$ and $\Delta S(\varphi - \pi/3)$ have the same distribution, which means that $C_{\Psi_3}$ and $C_{\Psi_3}^{\perp}$ have the same shape and $R_{\Psi_3}$ must be flat and have the value $1$.

This flat $R_{\Psi_3}$ can also be explained by the analysis based on CLT in Sec.~\ref{analyze}.


\item The $C_{\Psi_3}$ and $C_{\Psi_3}^\perp$ curves from Definition \textbf{A} show a similar dependence on resonance $p_{T,\rho}$ as $C_{\Psi_2}$ and $C_{\Psi_2}^\perp$ curves in the $v_2$ case.

\item The $C_{\Psi_3}$, $C_{\Psi_3}^{\perp}$, and $R_{\Psi_3}$ curves from Definition \textbf{B} are obviously dependent on the $p_{T,\rho}$ and $v_{3,\rho}$. 
Increasing $p_{T,\rho}$ makes the curves more convex. Increaing $v_{3,\rho}$ makes the $C_{\Psi_3}$, $R_{\Psi_3}$ curves more concave, and $C_{\Psi_3}^\perp$ more convex. Those tendencies are consistent with the scans with respect to $v_2$.

\item In Definition \textbf{B}, the $C_{\Psi_3}^\perp$ and $R_{\Psi_3}$ curves are counterintuitively not flat, even if we set $v_3$ to zero. 

\end{itemize}

We note that Definition \textbf{A} was used only in the early version (version~2) of Ref.~\cite{Magdy:2017yjev2} where the $R_{\Psi_3}(\Delta S)$ variable was studied with respect to $v_3$. In the later version~3 of Ref.~\cite{Magdy:2017yjev2}, Definition \textbf{B} was used.


\section{Analytical results based on the central limit theorem} \label{analyze}

In this section, we use the central limit theorem (CLT) to analyze the sine observable. This analysis can be applied to all observables discussed in this paper. With a few reasonable approximations, the behavior of the sine observable can be readily understood.

There are many versions of the CLT, and here we use the \emph{Lindeberg-Levy} expression. 
Let $X_1, X_2, \ldots, X_n$ be a sequence of independent and identically distributed (i.d.d.) random variables with expectation value $\text{E}[X_i] = \mu$ and variance $\text{Var}[X_i] = \sigma^2 < \infty$, and
\begin{equation}
S_n = \frac{X_1 + X_2 + \cdots + X_n}{n}
      = \frac{1}{n}\sum_{i=1}^{n} X_i
\end{equation}
denotes their mean. As $n$ approaches infinity, the random variable $\sqrt{n}(S_n - \mu)$ converges in distribution to a normal $\mathcal{N}(0, \sigma^2)$.
Generally, if $X_1, X_2, \ldots, X_n$ are independent normal distributions,
\begin{equation}
X_i \sim \mathcal{N}(\mu_i, \sigma_i^2)
,
\end{equation} 
then the weighted sum of them is a normal distribution,
\begin{equation}
\sum_{i=1}^{n} a_iX_i \sim \mathcal{N}\left(\sum_{i=1}^{n} a_i\mu_i, \sum_{i=1}^{n} a_i^2\sigma_i^2\right)
.
\end{equation}

\subsection{Analysis of $C_{\Psi_2}$, $C_{\Psi_2}^{\perp}$, and $R_{\Psi_2}$}

First, we write the 
PDF of $\phi$, 
\begin{equation} \label{PDF}
f(\phi)=\frac{1}{2\pi}\left(1+2\sum_{m=2}^{\infty} v_{m}\cos\left(m\left(\phi-\Psi_m\right)\right)\right)
,
\end{equation}
where $\Psi_m$ are normally different and uncorrelated among different $m$. 
When we focus only on one specific $m$, for example $m=2$ or $3$ in the former discussion, we can just use $\varphi=\phi-\Psi_m$ as the relative azimuth of particles.

\subsubsection{Numerator of $C_{\Psi_2}$} \label{CpNumerator}

The PDF of $\Delta S_{sep}$ can describe $N(\Delta S_{sep})$, the numerator of $C_{\Psi_2}$.
For simplicity, we assume that the number of positive charges is the same as the number of negative charges in the final state. In each event, before any decay, $n_\rho$ denotes the number of $\rho$ mesons, and $n_\pi$ denotes the number of primordial pions. Thus,
\begin{equation}
N_n = N_p = n_\rho + 0.5n_\pi
.
\end{equation}
We rewrite
\begin{equation}\label{Ssep}
\begin{split}
\Delta S_{sep} =& \frac{1}{n_\rho + 0.5n_\pi}\left(\sum_1^{n_\rho} (\sin\varphi_+ - \sin\varphi_-)\right)\\
&+ \frac{1}{n_\rho + 0.5n_\pi}\left(\sum_1^{0.5n_\pi} (\sin\varphi_+ - \sin\varphi_-) \right)
.
\end{split}
\end{equation}
The first sum is over $\rho$ decay pions, and the second is over primordial pions.

For convenience, we will use the following shorthand notations:
\begin{equation} \label{short}
\begin{split}
c:=\cos\varphi, \quad \bar{c}:=\cos\bar{\varphi}, \quad s:=\sin\varphi, \quad \bar{s}:=\sin\bar{\varphi},\\ 
\delta:=2\sin(\delta\varphi/2) 
\end{split}
\end{equation}
where $\bar{\varphi} = (\varphi_+ + \varphi_-)/2$ is related to the $\rho$ angular position and $\delta\varphi = \varphi_+ - \varphi_-$ represents the decay opening angle.
We use the indices $\rho$ or $\pi$ to indicate whether the variables are for $\rho$ or primordial $\pi^\pm$. 

We express the first sum of Eq.~\ref{Ssep} as
\begin{equation}
\sum_1^{n_\rho} (\sin\varphi_+ - \sin\varphi_-) = \sum_1^{n_\rho}2\cos(\bar{\varphi})\sin(\delta\varphi/2)
=\sum_{1}^{n_\rho}\bar{c}_\rho\delta
.
\end{equation}
Because the primordial pions all independently obey the same distribution related to the global harmonic plane, we rewrite the second sum of Eq.~\ref{Ssep} as
\begin{equation} \label{uuupi}
\sum_1^{0.5n_\pi} (\sin\varphi_+ - \sin\varphi_-) = 
\sum_1^{0.5n_\pi} \sin\varphi_+ - \sum_1^{0.5n_\pi} \sin\varphi_-
.
\end{equation}

We make two assumptions: 
(1) In a resonance decay, $\bar{\varphi}$ could be regarded as an approximation for $\varphi_\rho$, so the PDF of $\bar{\varphi}$ is the same as the PDF of $\varphi_\rho$.
(2) For two tracks from one resonance decay, $\cos\bar{\varphi}$ and $2\sin(\delta\varphi/2)$ are independent.

From symmetry, $\text{E}[\delta] = \text{E} [ 2\sin(\delta\varphi/2) ] = 0$ at any given $\bar{c}_\rho$, so
\begin{equation}
\text{E}[\bar{c}_\rho\delta]=\text{E}[\bar{c}_\rho]\text{E}[\delta]=\text{E}[\bar{c}_\rho]\times0=0
.
\end{equation}
We therefore get
\begin{equation}
\begin{split}
\text{Var}[\bar{c}_\rho\delta]
&=\text{Var}[\bar{c}_\rho]\text{Var}[\delta]+\text{E}[\bar{c}_\rho]^2 \text{Var}[\delta]+\text{Var}[\bar{c}_\rho]\text{E}[\delta]^2\\
&=\text{E}\left[\bar{c}_\rho^2\right] \text{Var}[\delta]
.
\end{split}
\end{equation}

In our simulations, $n_\rho$ is a Poisson distribution, so to get the variance of $\sum_{1}^{n_\rho}\bar{c}_\rho\delta$ is a problem of the compound Poisson distribution. Thus, we have
\begin{equation}\label{CompoundPoisson}
\begin{split}
\text{Var}\left[\sum_{i}^{n_\rho}\bar{c}_\rho\delta\right] &= \text{E}\left[n_\rho\right]\text{Var}[\bar{c}_\rho\delta]+\text{E}[\bar{c}_\rho\delta]^2\text{Var}[n_\rho]\\
&= \text{E}\left[n_\rho\right]\text{Var}[\bar{c}_\rho\delta]
.
\end{split}
\end{equation}
Equation~\ref{CompoundPoisson} indicates that it makes no difference whether $n_\rho$ is a single value or a Poisson distribution. For simplicity, we can just use $n_\rho$ as if it is fixed to a specific value. 
According to CLT,
\begin{equation}
\sum_{1}^{n_\rho}\bar{c}_\rho\delta \sim \mathcal{N} \left( 0, n_\rho\text{Var}[\delta]\text{E}\left[\bar{c}_\rho^2\right] \right) 
.
\end{equation}

The PDF of $\varphi$ of the primordial pions has the same form as Eq.~\ref{PDF}, so we can readily obtain the variances ($\text{Var}[c_\pi]$ and $\text{Var}[s_\pi]$).
As for the term about primordial pions, the two terms in the right hand side of Eq.~\ref{uuupi} should have the same distribution.
The discussion about $n_\pi$ is as same as the discussion of $n_\rho$. According to CLT, we have
\begin{equation}
\sum_1^{0.5n_\pi} \sin\varphi_+, \sum_1^{0.5n_\pi} \sin\varphi_- \sim \mathcal{N}(0.5n_\pi\text{E}[s_\pi], 0.5n_\pi\text{Var}[s_\pi])
,
\end{equation}
so the difference is
\begin{equation} \label{uuudif}
\sum_1^{0.5n_\pi} \sin\varphi_+ - \sum_1^{0.5n_\pi} \sin\varphi_- \sim \mathcal{N}(0,n_\pi \text{Var}[s_\pi])
.
\end{equation}

Finally, we write $\Delta S$ in our new notation, 
\begin{equation}
\Delta S_{sep} = 
\frac{\sum_{1}^{n_\rho}\bar{c}_\rho\delta + \sum_{1}^{n_\pi/2} s_{\pi^+} - \sum_{1}^{n_\pi/2} s_{\pi^-}}{n_\rho + 0.5n_\pi}
,
\end{equation}
where $s_{\pi^+}$ and $s_{\pi^-}$ are the sine values for $\pi^+$ and $\pi^-$ from resonance decay, and they obey the same distribution independently, so we just call them both $s_\pi$. According to CLT, 
\begin{equation}
\begin{split}
\Delta S_{sep} 
&\sim \mathcal{N} \left(0,\frac{n_\rho\text{Var}[\delta]\text{E}\left[\bar{c}_\rho^2\right]+n_\pi\text{Var}[s_\pi]}{(n_\rho + 0.5n_\pi)^2} \right)\\
&:= \mathcal{N} \left(0,\sigma_\uparrow^2 \right)
.
\end{split}
\end{equation}

\subsubsection{Denominator of $C_{\Psi_2}$}

The PDF of $\Delta S_{mix}$ can describe $N(\Delta S_{mix})$, the denominator of $C_{\Psi_2}$. The analysis here is very similar to the analysis of $\Delta S_{sep}$. In shuffling, we keep the number of positive charges still the same as the number of negative charges:
\begin{equation}
N_n' = N_p' = n_\rho + 0.5n_\pi
\end{equation}
Relaxing this requirement to an average level does not change our results.

After shuffling, all the pions are independent, no matter whether they are primordial or from resonance decays. For pions from resonance decays, the pion azimuth can be written as $\varphi=\bar{\varphi}+\delta\varphi/2 \approx \varphi_\rho+\delta\varphi/2$. Because the distribution of $\delta\varphi$ is symmetric about $\delta\varphi = 0$, we just use $+\delta\varphi/2$ here.
The expression of $\Delta S_{mix}$ can therefore be rewritten as:
\begin{equation}
\begin{split}
\Delta S_{mix} 
=& \frac{1}{n_\rho + 0.5n_\pi}\left(\sum_1^{n_\rho} (\sin\varphi_+ - \sin\varphi_-)\right) \\
&+ \frac{1}{n_\rho + 0.5n_\pi}\left(\sum_1^{0.5n_\pi} (\sin\varphi_+ - \sin\varphi_-) \right) \\
=& \frac{\sum_1^{n_\rho} \sin\left(\varphi_\rho+\delta\varphi_+/2\right) - \sum_1^{n_\rho}\sin\left(\varphi_\rho+\delta\varphi_-/2\right)}{n_\rho + 0.5n_\pi}\\
&+\frac{\sum_1^{n_\pi/2} s_{\pi^+} - \sum_1^{n_\pi/2} s_{\pi^-}}{n_\rho + 0.5n_\pi}
.
\end{split}
\end{equation}
The second term is already calculated in Eq.~\ref{uuudif}, and we calculate the distribution of the first term as
\begin{equation}
\begin{split}
&\sum_{1}^{n_\rho} \sin\left(\varphi_\rho+\frac{\delta\varphi_+}{2}\right) - \sum_{1}^{n_\rho} \sin\left(\varphi_\rho+\frac{\delta\varphi_-}{2}\right) \\
&\sim \mathcal{N} \left(0, 2n_\rho\text{Var}\left[\sin\left(\varphi_\rho+\frac{\delta\varphi}{2}\right)\right] \right),
\end{split}
\end{equation}
The first and the second moment below are needed in order to complete the calculation of the variance:
\begin{equation} \label{uuu1m}
\begin{split}
\text{E} &\left[\sin\left(\varphi_\rho + \frac{\delta\varphi}{2}\right)\right]\\
=& \text{E} \left[\sin\varphi_\rho\right] \text{E}\left[\cos\frac{\delta\varphi}{2}\right] + \text{E} \left[\cos\varphi_\rho\right] \text{E}\left[\sin\frac{\delta\varphi}{2}\right]\\
=& \text{E} \left[\sin\varphi_\rho\right] \text{E}\left[\cos\frac{\delta\varphi}{2}\right]
,
\end{split}
\end{equation}
\begin{equation} \label{uuu2m}
\begin{split}
\text{E} &\left[ \sin^2 \left(\varphi_\rho + \frac{\delta\varphi}{2}\right)\right]\\
=&\frac{1}{2}-\frac{1}{2}\text{E}\left[\cos(2\varphi_\rho)\cos(\delta\varphi)\right]+\frac{1}{2}\text{E}\left[\sin(2\varphi_\rho)\sin(\delta\varphi)\right]\\
=&\frac{1}{2}-\frac{1}{2}\text{E}\left[\left(1-2\sin^2\varphi_\rho\right)\left(1-\frac{1}{2}\delta^2\right)\right]+0\\
=&\text{E}[s_\rho^2]+\frac{1}{4}\text{Var}[\delta]-\frac{1}{2}\text{E}[s_\rho^2]\text{Var}[\delta]
.
\end{split}
\end{equation}
The last step uses the fact that $\text{E}[\delta]=0$ and therefore $\text{Var}[\delta]=\text{E}[\delta^2]$.

Thus, we can get the distribution of $\Delta S_{mix}$,
\begin{equation}
\begin{split}
\Delta S_{mix} &\sim \mathcal{N}\left(0, \frac{2n_\rho\text{Var}\left[\sin\left(\varphi_\rho+\frac{\delta\varphi}{2}\right)\right] + n_\pi\text{E}[s_\pi^2]}{(n_\rho + 0.5n_\pi)^2} \right) \\
&=: \mathcal{N} \left(0,\sigma_\downarrow^2 \right)
.
\end{split}
\end{equation}

\subsubsection{Shape of $C_{\Psi_2}$}

We use the PDF of a normal distribution Gaussian function:
\begin{equation}
f(x|\mu,\sigma) := \frac{1}{\sqrt{2\pi\sigma^2}}\exp{ \left(-\frac{(x-\mu)^2}{2\sigma^2} \right)}
.
\end{equation}
The shape of $C_{\Psi_2}$ is described by the ratio of the PDF of $\Delta S_{sep}$ to the PDF of $\Delta S_{mix}$. Using Gaussian functions for those PDFs, the shape of $C_{\Psi_2}$ is
\begin{equation}
C_{\Psi_2}(x)=\frac{f(x|0,\sigma_\uparrow)}{f(x|0,\sigma_\downarrow)} = \frac{\sigma_\downarrow}{\sigma_\uparrow}\exp{ \left[-\frac{x^2}{2} \left(\frac{1}{\sigma_\uparrow^2}-\frac{1}{\sigma_\downarrow^2} \right) \right]} 
.
\end{equation}
Here, $x$ denotes $\Delta S$ (representing $\Delta S_{sep}$ or $\Delta S_{mix}$).


\subsubsection{Shape of $C_{\Psi_2}^{\perp}$}

The analysis of $C_{\Psi_2}^{\perp}$ is nearly the same as that of $C_{\Psi_2}$ by shifting the relative azimuth $\varphi$ by a centain angle: $\varphi'=\varphi - \xi$. Accordingly, we use the parallel shorthand notations as follows:
\begin{equation}
\begin{split}
c':=\cos\varphi', \quad \bar{c}':=\cos\bar{\varphi}', \quad s':=\sin\varphi', \\
\bar{s}':=\sin\bar{\varphi}',\quad
\delta:=2\sin(\delta\varphi'/2)=2\sin(\delta\varphi/2)
.
\end{split}
\end{equation}
Then, the format of variances here is just like before:
\begin{equation}
\sigma_{\perp\uparrow}^2=\frac{n_\rho\text{Var}[\delta]\text{E}[\bar{c}_\rho'^2]+n_\pi\text{Var}[s_\pi']}{(n_\rho + 0.5n_\pi)^2} 
,
\end{equation}
\begin{equation}
\sigma_{\perp\downarrow}^2=\frac{2n_\rho\text{Var}\left[\sin\left(\varphi_\rho'+\frac{\delta\varphi}{2}\right)\right] + n_\pi\text{E}[s_\pi'^2]}{(n_\rho + 0.5n_\pi)^2}
.
\end{equation}
The shape of $C_{\Psi_2}^{\perp}$ is
\begin{equation}
C_{\Psi_2}^{\perp}(x)=
\frac{f(x|0,\sigma_{\perp\uparrow})}{f(x|0,\sigma_{\perp\downarrow})} = \frac{\sigma_{\perp\downarrow}}{\sigma_{\perp\uparrow}}\exp{ \left[-\frac{x^2}{2} \left(\frac{1}{\sigma_{\perp\uparrow}^2}-\frac{1}{\sigma_{\perp\downarrow}^2} \right) \right]} 
.
\end{equation}



\subsubsection{Shape of $R_{\Psi_2}$}
According to the definition of $R_{\Psi_2}$, the shape of $R_{\Psi_2}$ is given by
\begin{equation}
\begin{split}
R_{\Psi_2}(x)=&
\frac{f(x|0,\sigma_\uparrow)}{f(x|0,\sigma_\downarrow)} \left/ \frac{f(x|0,\sigma_{\perp\uparrow})}{f(x|0,\sigma_{\perp\downarrow})} \right. \\
=& \frac{\sigma_\downarrow\sigma_{\perp\uparrow}}{\sigma_\uparrow\sigma_{\perp\downarrow}}\exp{ \left[-\frac{x^2}{2} \left(\frac{1}{\sigma_\uparrow^2}-\frac{1}{\sigma_\downarrow^2} - \frac{1}{\sigma_{\perp\uparrow}^2}+\frac{1}{\sigma_{\perp\downarrow}^2} \right) \right]}
.
\end{split}
\end{equation}
Thus, whether $R_{\Psi_2}$ is convex or concave is determined by the following parameter:
\begin{equation}
\zeta := \frac{1}{\sigma_\uparrow^2}-\frac{1}{\sigma_\downarrow^2} - \frac{1}{\sigma_{\perp\uparrow}^2}+\frac{1}{\sigma_{\perp\downarrow}^2}
.
\end{equation}

\begin{itemize}
\item If $\zeta > 0$, then $R_{\Psi_2}$ is convex, and the more positive $\zeta$ is, the more convex $R_{\Psi_2}$ will be.
\item If $\zeta < 0$, then $R_{\Psi_2}$ is concave, and the more negative $\zeta$ is, the more concave $R_{\Psi_2}$ will be.
\item If $\zeta = 0$, then $R_{\Psi_2}$ is flat.
\end{itemize}

\subsection{CLT analysis for $v_2$}

If we only focus on $v_2$, the PDF in Eq.~\ref{PDF} can be simplified as 
\begin{equation}
f(\varphi)=\frac{1}{2\pi}\left(1+2v_2\cos(2\varphi)\right)
.
\end{equation}
From the definition of $C_{\Psi_2}^{\perp}$ for $v_2$, the relative azimuth is shifted by $\xi=\pi/2$. Thus, 
\begin{equation} \label{uuuCpm2}
\begin{split}
&\text{E}[c_\rho^2]=\text{E}[\bar{c}_\rho^2]=\frac{1+v_{2,\rho}}{2},
\quad \text{E}[s_\rho^2]=\frac{1-v_{2,\rho}}{2},\\
&\text{E}[c_\rho'^2]=\text{E}[\bar{c}_\rho'^2]=\frac{1-v_{2,\rho}}{2},
\quad \text{E}[s_\rho'^2]=\frac{1+v_{2,\rho}}{2},\\
&\text{E}[s_\pi^2]=\frac{1-v_{2,\pi}}{2},
\quad \text{E}[s_\pi'^2]=\frac{1+v_{2,\pi}}{2}
.
\end{split}
\end{equation}
We can easily get that the first moment of $\sin\left(\varphi_\rho + \delta\varphi/2\right)$ in Eq.~\ref{uuu1m} is $0$, so its variance is equal to its second moment which can be expressed as Eq.~\ref{uuu2m} by the terms in Eq.~\ref{uuuCpm2}.


After slightly changing the sequence in the expression of $\zeta$, we have 
\begin{equation}
\begin{split}\label{zeta2}
\zeta
=& \left( \frac{1}{\sigma_\uparrow^2} - \frac{1}{\sigma_{\perp\uparrow}^2} \right) - \left( \frac{1}{\sigma_\downarrow^2} - \frac{1}{\sigma_{\perp\downarrow}^2} \right)\\
=& \bigg( \frac{2(n_\rho + 0.5n_\pi)^2}{n_\rho\text{Var}[\delta](1+v_{2,\rho})+ n_\pi(1-v_{2,\pi})} \\
&- \frac{2(n_\rho + 0.5n_\pi)^2}{n_\rho\text{Var}[\delta](1-v_{2,\rho}) + n_\pi(1+v_{2,\pi})} \bigg)\\
&- \bigg( \frac{2(n_\rho + 0.5n_\pi)^2}{2n_\rho(1-v_{2,\rho}) + n_\rho v_{2,\rho} \text{Var}[\delta] + n_\pi(1-v_{2,\pi})} \\
&- \frac{2(n_\rho + 0.5n_\pi)^2}{2n_\rho(1+v_{2,\rho}) -  n_\rho v_{2,\rho} \text{Var}[\delta] + n_\pi(1+v_{2,\pi})} \bigg)
.
\end{split}
\end{equation}

For further insights, we make two more assumptions (in addition to those in Sec.~\ref{CpNumerator}):
(3) The magnitude of $v_2$ (including $v_{2,\rho}$ and $v_{2,\pi}$) is much smaller than 1. In our simulations, they are around 0.1;
(4) In each event, the number of primordial pions are much larger than the number of $\rho$ mesons. In our simulations, $n_\pi \approx 10n_\rho$.

In our simulations, $v_{2,\rho}$, $v_{2,\pi}$, and $n_\rho/n_\pi$ are of the same order of magnitude ($\sim 0.1$). To the leading order of them,
\begin{equation}
\zeta = \frac{n_\rho}{n_\pi} (2n_\rho + n_\pi)^2 \left(\frac{ 4v_{2,\pi}-2v_{2,\rho}-2v_{2,\pi}\text{Var}[\delta]}{n_\pi+n_\rho \left(4+2\text{Var}[\delta]\right)}\right)
.
\end{equation}
The first derivatives are
\begin{equation}
\frac{\partial \zeta}{\partial v_{2,\rho}} = \frac{n_\rho}{n_\pi} (2n_\rho + n_\pi)^2 \left(\frac{-2}{n_\pi+n_\rho \left(4+2\text{Var}[\delta]\right)}\right) < 0
,
\end{equation}
\begin{equation}\label{zetavpi}
\frac{\partial \zeta}{\partial v_{2,\pi}} = \frac{n_\rho}{n_\pi} (2n_\rho + n_\pi)^2 \left(\frac{ 4-2\text{Var}[\delta]}{n_\pi+n_\rho \left(4+2\text{Var}[\delta]\right)}\right)
,
\end{equation}
\begin{equation}\label{zetavardelta}
\begin{split}
\frac{\partial \zeta}{\partial \text{Var}[\delta]} = \frac{-2n_\rho}{n_\pi}  \left(\frac{2n_\rho + n_\pi}{n_\pi+n_\rho \left(4+2\text{Var}[\delta]\right)}\right)^2\\
\times\left(v_{2,\pi}n_\pi+8v_{2,\pi}n_\rho-2v_{2,\rho} n_\rho \right)
.
\end{split}
\end{equation}
When $0 \le \text{Var}[\delta] < 2$, $\partial \zeta / \partial v_{2,\pi} >0$.
When $2 < \text{Var}[\delta] \le 4$, $\partial \zeta / \partial v_{2,\pi} <0$. 

Varying the $p_{T,\rho}$ changes $\text{Var}[\delta]$. As long as $v_{2,\rho}$ is no more than $9v_{2,\pi}$ (which is almost always the case), then $\partial \zeta / \partial \text{Var}[\delta] < 0$. In our $p_T$ scan, $v_{2,\rho}$ is a single value $0.06$, and the average of $v_{2,\pi}$ is also around this value. 

\begin{figure}[h]
\subfloat{
  \centering
  \includegraphics[width=0.9\linewidth]{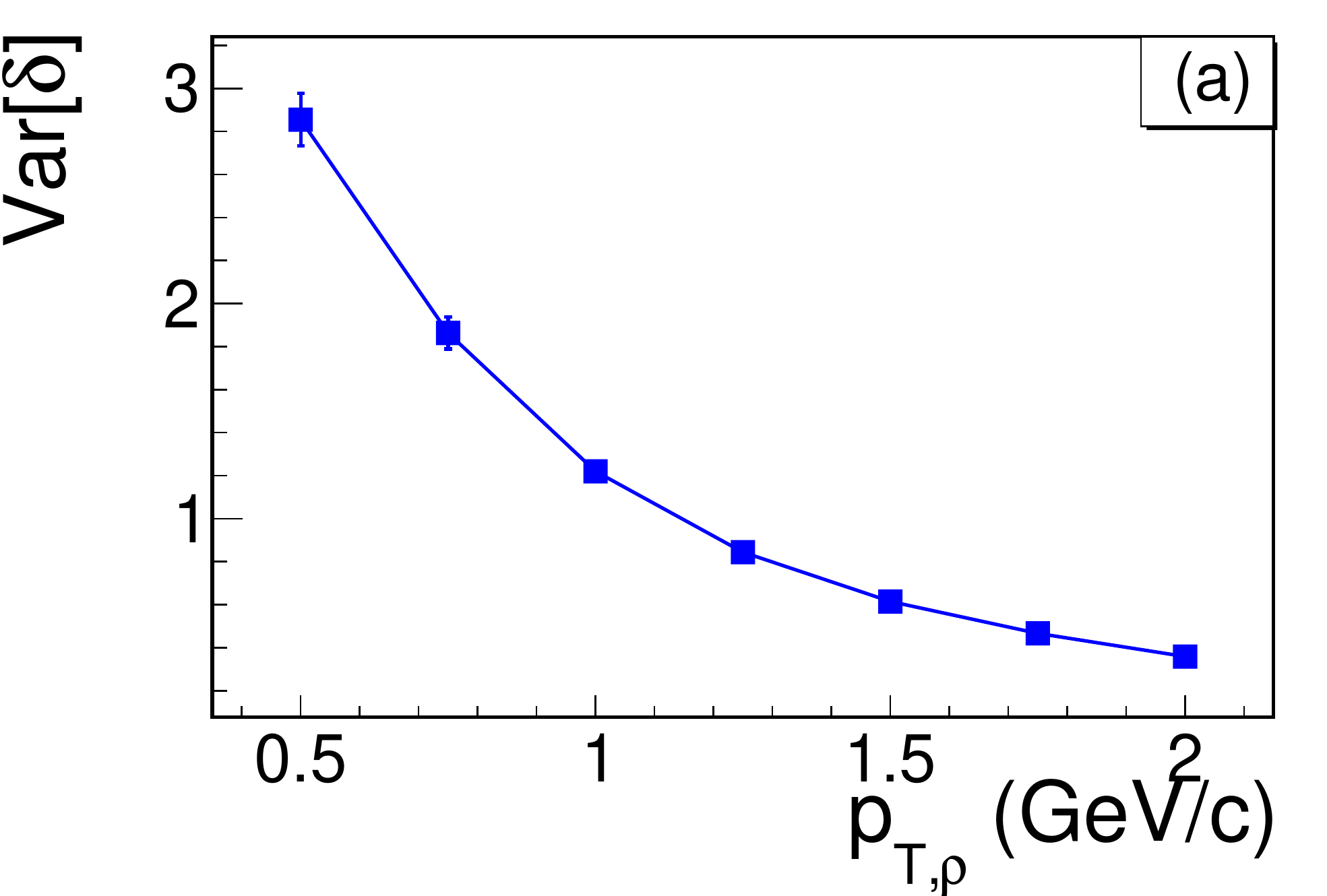}
  \label{Var-pT}
}

\centering
\subfloat{
  \centering
  \includegraphics[width=0.9\linewidth]{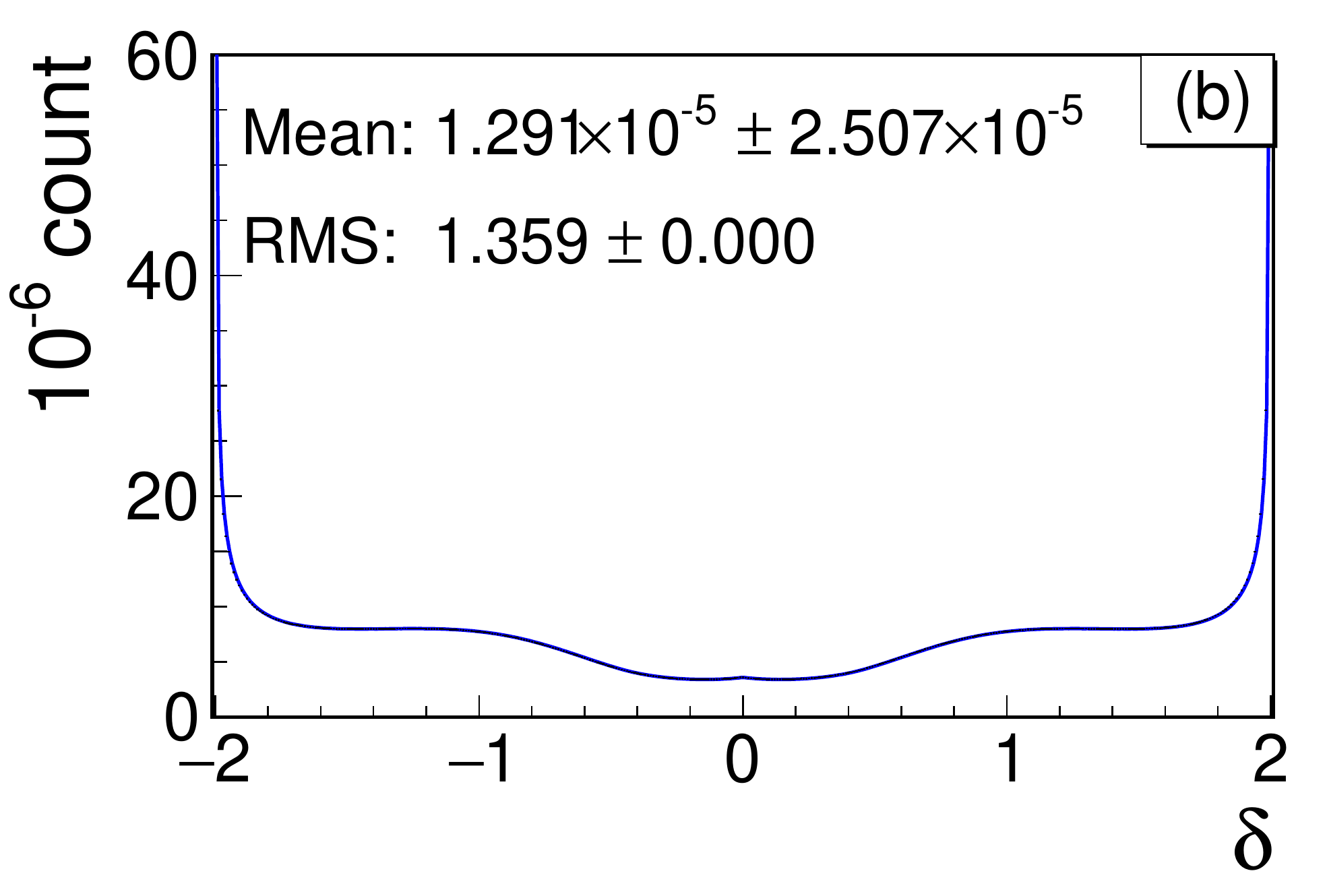}
  \label{deltaRMS}
}

\caption{$\text{Var}[\delta]$ in the default simulation and the scan of $p_{T,\rho}$. (a) $\text{Var}[\delta]$ depending on $p_{T,\rho}$. (b) $\delta$ distribution in the default simulation.}
\end{figure}

Thus, after suitable approximations, we can see the effects on the shape of $R_{\Psi_2}$ from those variables:
\begin{itemize}
\item Increasing $v_{2,\rho}$ makes $R_{\Psi_2}$ more concave.
\item Increasing $\text{Var}[\delta]$ makes $R_{\Psi_2}$ more concave. 
Increasing $p_{T,\rho}$ makes $R_{\Psi_2}$ more convex, because larger $p_{T,\rho}$ makes the two daughter pions closer to each other in angle, yielding a smaller $\text{Var}[\delta]$ (see Fig.~\ref{Var-pT}).
\item Increasing $v_{2,\pi}$ makes $R_{\Psi_2}$ more convex when $\text{Var}[\delta]<2$, and more concave when $\text{Var}[\delta]>2$. In our default simulation (Fig.~\ref{deltaRMS}), $\text{Var}[\delta] \approx 1.359^2 < 2$, so $R_{\Psi_2}$ becomes more convex as $v_{2,\pi}$ increases.
\end{itemize}

The conclusions of the CLT analysis are consistent with the simulation results.

\subsection{CLT analysis for $v_3$} \label{CLTv3}

If we only focus on $v_3$, the PDF in Eq.~\ref{PDF} could be simplified as follows:
\begin{equation}\label{v2PDF}
f(\varphi)=\frac{1}{2\pi}\left(1+2v_3\cos(3\varphi)\right)
.
\end{equation}

\subsubsection{Analysis for Definition \textbf{A}}
By Definition \textbf{A}, we list the shorthand notations:
\begin{equation}
\begin{split}
c:=\cos\varphi, \quad s:=\sin\varphi, \quad \xi=\pi/3, \\
c':=\cos\left(\varphi-\frac{\pi}{3}\right), \quad s':=\sin\left(\varphi-\frac{\pi}{3}\right), \\
\delta:=2\sin\left(\frac{1}{2}\delta\varphi\right)
.
\end{split}
\end{equation}
By using the simplified PDF (Eq.~\ref{v2PDF}), we can easily get the second moments needed:
\begin{equation}
\begin{split}
&\text{E}[{c}_\rho^2]=\text{E}[\bar{c}_\rho^2]=\frac{1}{2},
\quad \text{E}[s_\rho^2]=\frac{1}{2},
\quad \text{E}[s_\pi^2]=\frac{1}{2},\\
&\text{E}[c_\rho'^2]=\text{E}[\bar{c}_\rho'^2]=\frac{1}{2},
\quad \text{E}[s_\rho'^2]=\frac{1}{2},
\quad \text{E}[s_\pi'^2]=\frac{1}{2}.
\end{split}
\end{equation}
There is no $v_{3,\rho}$ or $v_{3,\pi}$ in any term above, so the shapes of the observables should not change with $v_{3,\rho}$ or $v_{3,\pi}$. We can just utilize the CLT analysis results for $v_2$ by setting all $v_2$ values to $0$, and then from the expression of $\zeta$ in Eq.~\ref{zeta2}, we see the terms in each bracket cancel each other. Thus, the CLT analysis shows $\zeta=0$, and accordingly, $R_{\Psi_3}$ should be always flat, as indeed shown in Fig.~\ref{N3R}.

\subsubsection{Analysis for Definition \textbf{B}}
\begin{figure}[h]
  \centering
  \includegraphics[width=0.33\textwidth]{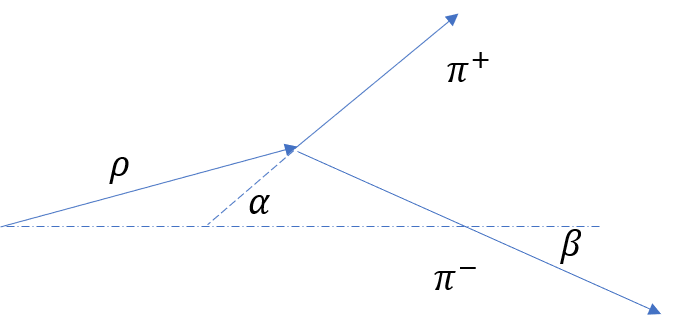}
  \caption{The choice of the azimuthal range affects the physical results using Definition \textbf{B}.}
  \label{ChargeDisorder}
\end{figure}

By Definition \textbf{B}, we list the shorthand notations:
\begin{equation}
\begin{split}
c:=\cos\left(\frac{3}{2}\varphi\right) \quad s:=\sin\left(\frac{3}{2}\varphi\right), \quad \xi=\frac{\pi}{3}, \\
\quad c':=\cos\left(\frac{3}{2}\left(\varphi-\frac{\pi}{3}\right)\right),
\quad s':=\sin\left(\frac{3}{2}\left(\varphi-\frac{\pi}{3}\right)\right),\\
\delta :=2\sin\left(\frac{3}{4}\delta\varphi\right)
.
\end{split}
\end{equation}
From the simplified PDF (Eq.~\ref{v2PDF}), we can get the first and the second moments:
\begin{equation} \label{R3m1}
\begin{split}
&\text{E}[s_\rho] = \text{E}[c_\rho'] = \frac{6-4v_{3,\rho}}{9\pi}, \quad \text{E}[c_\rho] = \text{E}[s_\rho'] = 0,\\
&\text{E}[s_\pi] = \text{E}[c_\pi'] = \frac{6-4v_{3,\pi}}{9\pi}, \quad \text{E}[c_\pi] = \text{E}[s_\pi'] = 0
,
\end{split}
\end{equation}
\begin{equation}
\begin{split}
&\text{E}[c_\rho^2]=\text{E}[\bar{c}_\rho^2]=\frac{1+v_{3,\rho}}{2},
\quad \text{E}[s_\rho^2]=\frac{1-v_{3,\rho}}{2},\\
&\text{E}[c_\rho'^2]=\text{E}[\bar{c}_\rho'^2]=\frac{1-v_{3,\rho}}{2},
\quad \text{E}[s_\rho'^2]=\frac{1+v_{3,\rho}}{2},\\
&\text{E}[s_\pi^2]=\frac{1}{2},
\quad \text{E}[s_\pi'^2]=\frac{1}{2}
,
\end{split}
\end{equation}
where we have a constraint that the azimuthal range must be $0 \le \varphi < 2\pi$. 
Because of the non-zero first moments, the $R_{\Psi_3}$ curve is not flat ($\zeta \neq 0$) even if both $v_{3,\rho}$ and $v_{3,\pi}$ are set to $0$. This counterintuitive observation is due to the absence of the periodical symmetry in the Definition \textbf{B}. For the same reason, Definition \textbf{B} has some disadvantages as follow:
\begin{itemize}
\item The $R_{\Psi_3}$ curve is counterintuitively not flat, even if both $v_{3,\rho}$ and $v_{3,\pi}$ are set to $0$ which means all azimuths are isotropically distributed.
\item The azimuthal range must be set. In the former discussion, we let $\varphi \in [0,2\pi)$. However, if we let the azimuthal range be $\varphi \in [-\pi,\pi)$, the first moments will change from Eq.~\ref{R3m1} into
\begin{equation}
\begin{split}
&\text{E}[s_\rho] = \text{E}[c_\rho'] = 0, \quad \text{E}[c_\rho] = \text{E}[s_\rho'] = \frac{6-4v_{3,\rho}}{9\pi},\\
&\text{E}[s_\pi] = \text{E}[c_\pi'] = 0, \quad \text{E}[c_\pi] = \text{E}[s_\pi'] = \frac{6-4v_{3,\rho}}{9\pi}
,
\end{split}
\end{equation}
which can make obvious differences to the features of the sine observables.

\item The azimuthal range is by choice, however, it introduces artificial unphysical differences using Definition \textbf{B}.
Take Fig.~\ref{ChargeDisorder} as an example. If we take the azimuthal range $[-\pi,\pi)$,
we have $\alpha > 0$ and $\beta < 0$. However, if we take the range $[0,2\pi)$, $\beta$ will become $\beta' = \beta + 2\pi$. The contribution of this resonance decay to $\Delta S_{sep}$ changes from 
\[\quad\quad\quad\sin\left(\frac{3}{2}\alpha\right) - \sin\left(\frac{3}{2}\beta\right)\]
into
\[\quad\quad\quad\sin\left(\frac{3}{2}\alpha\right) - \sin\left(\frac{3}{2}\beta'\right) = \sin\left(\frac{3}{2}\alpha\right) + \sin\left(\frac{3}{2}\beta\right).\]
It seems just like the negative charge becomes a positive one.
\end{itemize}

We thus conclude that Definition \textbf{B} is ill-devised, and should not be used. On the other hand, Definition \textbf{A} always yields a flat $R_{\Psi_3}$ distribution and therefore is not sensitive to the CME or background. It therefore appears that the $\Psi_3$ harmonic plane is not suitable for the sine observables.

\section*{Summary}
We have presented a systematic study of resonance backgrounds as functions of the resonance $v_2$ and $p_T$ with toy-model simulations and CLT calculations, in order to better understand the behaviors of the sine observable.
It is found that the concavity or convexity of $R_{\Psi_2}(\Delta S)$ depends sensitively on the resonance $v_2$ 
(which yields different numbers of decay $\pi^+\pi^-$ pairs in the in-plane and out-of-plane directions) 
and $p_T$ (which affects the opening angle of the decay $\pi^+\pi^-$ pair).
Qualitatively, low $p_{T}$ resonances decay into large opening-angle pairs and result in more ``back-to-back'' pairs out-of-plane (because of the positive resonance $v_2$), mimicking a CME signal, or a concave $R_{\Psi_2}(\Delta S)$. High $p_T$ resonances, on the other hand, result in more close pairs in-plane, constituting a well-known background, or convex $R_{\Psi_2}(\Delta S)$. In other words, resonance backgrounds can yield both concave and convex $R_{\Psi_2}(\Delta S)$ distributions, depending on the resonance kinematics. 

We have also conducted a supplemental study using the triangular flow ($v_3$) and discussed two definitions for the sine variables. For one of the definitions, it is found that $R_{\Psi_3}(\Delta S)$ is
always flat due to the inherited symmetry in the definition. 
For the other definition, $R_{\Psi_3}(\Delta S)$ for $v_3$ is found to to behave similarly as $R_{\Psi_2}(\Delta S)$ for $v_2$, if the azimuthal angle is kept in the range $[0,2\pi)$; 
$R_{\Psi_3}$ can be concave or convex depending on details.
However, $R_{\Psi_3}$ is found to depend on the choice of the azimuthal angle range due to the inconsistency between the periods of $R_{\Psi_3}$ ($4\pi/3$) and azimuthal position ($2\pi$). If $[-\pi,\pi)$ is chosen to be the range, then the $R_{\Psi_3}$ results are completely different. 
Therefore, the $\Psi_3$ may not be suitable for the sine-observable studies. One has to be careful to keep the identical azimuthal angle range in the model-data comparison studies.

We have verified our toy-model simulation results by analytical CLT calculations.

If the CME is the only source for the RP-dependent and charge-dependent correlations, then the $R_{\Psi_2}(\Delta S)$ would be concave and $R_{\Psi_3}(\Delta S)$ would be convex for the nontrivial defintion. However, given the existence of backgrounds, a concave $R_{\Psi_2}(\Delta S)$ and a simultaneous convex $R_{\Psi_3}(\Delta S)$ do not lead to the conclusion of CME. 
This is because the $R_{\Psi_2}(\Delta S)$ and $R_{\Psi_3}(\Delta S)$ variables do not necessarily have a prior relationship, each individually varying with their respective $v_m(p_T)$ $(m=2,3)$ of resonances, and because the $R_{\Psi_3}$ variable depends on what azimuthal range is used.
Based on our results, it is clear that the qualitative concavity or convexity of the $R_{\Psi_2}(\Delta S)$ or $R_{\Psi_3}(\Delta S)$ variable, or the comparison between them, cannot conclude on the existence, nor the magnitude, of the CME. Since the $R_{\Psi_2}(\Delta S)$ and  $R_{\Psi_3}(\Delta S)$ variables depend on the details of the resonance kinematics and anisotropies, as well as the resonance abundances, a precise knowledge of all resonance distributions is required in order to quantify the CME using the $R_{\Psi_m}(\Delta S)$ $(m=2,3)$ observables. 

\section*{Acknowledgments}

Y.~Feng thanks Dr.~Wendell Lutz and Mrs.~Nancy Lutz~for their generous support of the Rolf Scharenberg Graduate Research Fellowship. 
We thank Roy Lacey and Niseem Magdy for useful discussions. This work is supported in part by the U.S.~Department of Energy Grant No.~DE-SC0012910 and the National Natural Science Foundation of China Grants No.~11647306 and No.~11747312.


\bibliography{../ref}

\end{document}